\documentclass{aa}
\usepackage{natbib}
\usepackage{graphicx}
\usepackage{amsmath}
\usepackage{amssymb}
\usepackage{latexsym}
\usepackage{lscape}
\usepackage{booktabs}
\usepackage{dcolumn}
\newcolumntype{d}[1]{D{.}{.}{#1}}
\newcolumntype{p}[1]{D{-}{-}{#1}}
\renewcommand{\j}[2]{\mbox{J $= #1\rightarrow #2$}}
\newcommand{\ceighteeno}{\mbox{C$^{18}$O}}
\newcommand{\cseventeeno}{\mbox{C$^{17}$O}}

\begin{document}

\title{The Circumstellar Environment of High-Mass Protostellar
Objects.\\ IV. \cseventeeno\ Observations and Depletion}

\author{H.~S.~Thomas
  \and G.~A.~Fuller}
\institute{Jodrell Bank Centre for Astrophysics, Alan Turing Building, The University of Manchester, Manchester, M13 9PL,  United Kingdom}

\offprints{H.~S.~Thomas,
  \email{H.Thomas@postgrad.manchester.ac.uk}}
\titlerunning{HMPOs: Depletion}
\date{Received date / Accepted date}

\abstract
  {}
  { The presence of depletion (freeze-out) of CO around low-mass protostars is
    well established. Here we observe 84 candidate young high-mass sources in
    the rare isotopologues C$^{17}$O and C$^{18}$O to investigate whether
    there is evidence for depletion towards these objects.}
  {Observations of the \j{2}{1}\ transitions of \ceighteeno\ and
    \cseventeeno\ are used to derive the column densities of gas
    towards the sources and these are compared with those derived from
    submillimetre continuum observations. The derived fractional
    abundance suggests that the CO species show a
    range of degrees of depletion towards the objects. We then use the
    radiative transfer code RATRAN to model a selection of the sources
    to confirm that the spread of abundances is not a result of assumptions
    made when calculating the column densities.}
  {We find a range of abundances of \cseventeeno\ that cannot be accounted for
    by global variations in either the temperature or dust properties and so
    must reflect source to source variations. The most likely explanation is
    that different sources show different degrees of depletion of the CO.
    Comparison of the C$^{17}$O linewidths of our sources with those of CS
    presented by other authors reveal a division of the sources into two
    groups. { Sources with a CS linewidth $>3$\,km\,s$^{-1}$ have low abundances of
      \cseventeeno\ while sources with narrower CS lines have typically higher
      \cseventeeno\ abundances.}  We suggest that this represents an
    evolutionary trend. Depletion towards these objects shows that the gas
    remains cold and dense for long enough for the trace species to deplete.
    The range of depletion measured suggests that these objects have lifetimes
    of $2-4\times10^5$ years. }
   {}
 
\maketitle

\section{Introduction}
\label{sec:introduction}

Many questions regarding the formation mechanism of high-mass stars
remain unanswered. Young high-mass stars form within massive cores
and display features common to low-mass star formation such as
outflows and in some cases disks
\citep[e.g.][]{shepherd2005,cesaroni2007}. However the different
stages through which a core forming a high-mass star evolves remains
unclear.

In the cold, dense conditions within cores prior to their collapse,
trace molecules freeze-out onto the dust grains forming icy mantles.
This process reduces the abundances of species in the gas phase until
the grains are heated, desorbing the molecules off the grains and 
returning them back into the gas. This freeze-out and evaporation
cycle is potentially an important indicator of the age of the
material in star forming cores and how the material has evolved.

The transitions of the two rare isotopologues of CO, \ceighteeno\ and
\cseventeeno\ which have abundances  with respect to H$_2$  of
$\sim 2 \times 10^{-7}$ and $\sim 5 \times 10^{-8}$ respectively
\citep*{frerking82}, are often employed to probe the inner regions of
dense cores \citep{tafalla02,caselli99,redman02} as their transitions
are typically optically thin, especially those of \cseventeeno.  A
number of studies of various, but mainly low-mass, star forming
regions using these species have provided evidence for depletion of CO
towards the centres of cores
\citep{caselli99,kramer99,tafalla02,willacy98,savva03} with the degree
of depletion varying from from factors of a few to over an order of
magnitude compared to the canonical CO abundance.

In this paper we present new observations of candidate high-mass star
forming regions in C$^{17}$O and C$^{18}$O made in order to look for
evidence of depletion in these regions.

\section{Observations of C$^{17}$O and  C$^{18}$O}
\label{sec:observations}
\citet[hereafter SBSMW]{sridharan02} searched the IRAS catalogue and
identified 69 point sources which represent potentially massive, deeply
embedded protostars in the Galactic plane.  These sources have been studied
extensively by \citet{beuther02a,beuther02b} and have been mapped with SCUBA,
the James Clerk Maxwell Telescope (JCMT) bolometer array, at 850\,$\mu$m and
450\,$\mu$m by \citet*[hereafter WFS04]{wfs04} as well as being surveyed for
evidence of infall by \cite{fuller05}.  In many cases the submillimetre and
millimetre continuum maps show more than a single peak indicating multiple
sites of potential star formation. In total 112 850\,$\mu$m peaks were
identified of which we have observed 84 in C$^{17}$O.

We observed the \j{2}{1} rotational transitions of C$^{18}$O (219.560
\,GHz) and C$^{17}$O (224.714\,GHz) towards 31 sources during May 2004 at
the JCMT\footnote{The James Clerk Maxwell Telescope is operated
  by The Joint Astronomy Centre on behalf of the Particle Physics and
  Astronomy Research Council of the United Kingdom, the Netherlands
  Organisation for Scientific Research, and the National Research
  Council of Canada}  in Hawaii as part of observing program M04AU47.
We used the RxA3 receiver with the DAS autocorrelator with channels of
78\,kHz in position switching mode using a reference position of
(+600$''$, +600$''$).  The reference positions were chosen to be free
of \ceighteeno\ emission having been checked and used for previous
molecular observations. The pointing was checked every couple of
hours. The typical zenith opacity at 225\,GHz was between 0.15 and 0.4
and we achieved a typical rms of $\lesssim$0.25\,K. The data were baseline
subtracted using the SPECX package following normal routines.

As we are primarily interested in deriving column densities, in
further observations we chose to concentrate on the C$^{17}$O line
which has lower optical depth. A further 53 sources were observed in
C$^{17}$O \j{2}{1} in July and August 2005 (in observing program
M05BU46). In addition we observed a selection of these sources in the
\j{3}{2} transition of \cseventeeno\ (337.061\,GHz) with the RxB3
receiver. This time we achieved a typical rms of $\lesssim$0.15\,K.
These data were processed in the same way as the earlier data.

As the telescope beam is not perfectly coupled to the source it is
necessary to apply a correction factor to convert antenna temperature
into a main beam temperature where
\begin{displaymath}
T_{mb}=\frac{T_A^*}{\eta_{mb}}
\end{displaymath}
 
At the JCMT the main beam efficiencies are 0.69 at 220\,GHz with a beam size of
20$''$ and 0.63 at 337\,GHz with a beam size of 13$''$. The positions and dates of
observations for all the sources are given in Table \ref{tab:sourcelist}.

\section{Analysis}
\label{sec:results}

\linespread{1}
\begin{table*}
\begin{scriptsize}
  \begin{center}
 \begin{tabular}{ccccccccc}
  \toprule
      \toprule

     \multicolumn{1}{c}{}      &       
     \multicolumn{4}{c}{C$^{18}$O \j{2}{1}}            &
     \multicolumn{4}{c}{C$^{17}$O \j{2}{1}}       \\
     \cmidrule(r){2-5}
     \cmidrule(l){6-9}

     \multicolumn{1}{c}{WFS}   &
      \multicolumn{1}{c}{Peak $T_{mb}$}             &              
      \multicolumn{1}{c}{$v_{lsr}$}                &
      \multicolumn{1}{c}{$\Delta v$}                &
      \multicolumn{1}{c}{$\int T_{mb}dv$}     &
      \multicolumn{1}{c}{Peak $T_{mb}$}             &              
      \multicolumn{1}{c}{$v_{lsr}$} &
      \multicolumn{1}{c}{$\Delta v$}                &
      \multicolumn{1}{c}{$\int T_{mb}dv$}     \\

      \multicolumn{1}{c}{}  &
      \multicolumn{1}{c}{(K)}  &              
      \multicolumn{1}{c}{(km\,s$^{-1}$)}       &
      \multicolumn{1}{c}{(km\,s$^{-1}$)}     &
      \multicolumn{1}{c}{(K\,km\,s$^{-1}$)}   &
      \multicolumn{1}{c}{(K)}  & 
      \multicolumn{1}{c}{(km\,s$^{-1}$)}        &
      \multicolumn{1}{c}{(km\,s$^{-1}$)}      &
      \multicolumn{1}{c}{(K\,km\,s$^{-1}$)}   \\

      \midrule

 14   & 9.77    &33.26  &2.42 &25.61   & 2.59  & 33.22 &  2.45  &7.75    \\      
 16   & 10.54   &59.22  &3.36 &37.38   & 3.97  & 59.19 &  3.00  &14.03        \\     
 17   & 10.74   &45.08  &2.72 &32.74   & 3.45  & 45.09 &  2.67  &11.65     \\     
 20   & 3.30    &34.39  &2.82 &9.80    & 1.26  & 34.28 &  2.71  &4.19   \\      
 21   & 5.19    &34.39  &2.25 &12.74   & 1.75  & 34.35 &  2.29  &5.22    \\     
 22   & 8.99    &84.51  &2.23 &21.42   & 3.09  & 84.48 &  2.01  &8.01    \\     
 25   & 7.17    &77.85  &2.47 &19.64   & 2.52  & 77.78 &  2.25  &7.62  \\       
 29   & 5.48    &58.94  &2.69 &15.22   & 2.67  & 59.24 &  2.16  &7.59    \\     
 30   & 4.06    &95.69  &2.44 &13.01   & 0.96  & 95.75 &  2.87  &3.70    \\     
 34   & 8.09    &22.83  &3.39 &30.06   & 2.62  & 22.96 &  3.41  &10.97      \\    
 36   & 4.67    &15.79  &2.64 &16.12   & 1.99  & 15.93 &  2.58  &6.39     \\    
 39   & 5.13    &96.07  &2.64 &17.28   & 1.91  & 96.04 &  2.63  &6.46     \\    
 51   & 7.75    &84.29  &1.85 &14.84   & 2.32  & 84.33 &  1.75  &5.16     \\    
 78   & 4.99    &21.72  &2.83 &15.12   & 1.32  & 21.80 &  2.47  &4.22     \\    
 79   & 6.01    &22.53  &2.85 &19.07   & 1.74  & 22.54 &  2.34  &6.10     \\    
 85   & -       &-  &-  &-             & 1.41  & 11.20 &  2.40  &4.45      \\    
 87   & -       &-  &-  &-             & 2.64  & 5.58  &  1.26  &5.04      \\   
 88   & -       &-  &-  &-             & 4.25  & 5.83  &  0.93  &6.43   \\
 90   & 5.43    &-3.64  &2.83 &17.29   & 1.46  & -3.67 &  2.92  &5.54      \\   
 91   & 10.38   &-1.89  &1.26 &14.01   & 3.13  & -1.85 &  1.20  &5.78        \\  
 95   & -       &-  &-  &-             & 0.94  & 8.52  &  2.33  &2.80       \\  
 96   & -       &-  &-  &-             & 1.87  & -3.19 &  2.01  &4.89            \\  
 97   & -       &-  &-  &-             & 1.33  & -2.19 &  2.77  &4.36              \\  
 99   & 6.01    &11.16  &2.10  &13.46  & 1.80  & 11.12 &  1.80  &4.23        \\  
 100  & 4.94    &11.40  &2.85  &14.12  & 1.54  & 11.40 &  2.60  &4.81        \\  
 107  & 4.09    &-45.85 &2.25  &10.38  & 1.26  & -45.86&  1.85  &3.01       \\ 
 108  & 5.10    &-53.18 &2.79  &16.48  & 1.52  & -53.11&  2.82  &5.88       \\ 
 109  & 4.36    &-44.29 &2.74  &13.47  & 1.29  & -44.38&  2.33  &4.00           \\ 
 110  & 3.91    &-54.71 &2.31  &10.09  & 1.22  & -54.73&  2.39  &3.62      \\ 
 111  & 10.17   &-17.94 &1.46  &15.81  & 3.29  & -17.88&  1.26  &6.28      \\ 
 112  & 6.68    &-18.45 &1.84  &12.51  & 1.87  & -18.62&  1.55  &4.01    \\   
  
\hline

\end{tabular}
\end{center}
\caption{\footnotesize Observed line parameters. The columns give the peak intensity,
velocity, linewidth and integrated intensity for \ceighteeno\  and
\cseventeeno\  respectively. For \cseventeeno\ the first three parameters are derived from hyperfine fitting  of the data  using CLASS. The integrated intensity was obtained using a
  single Gaussian fit.}  
\label{tab:tablehfs}
\end{scriptsize}

\end{table*}

\linespread{1.4}  
\linespread{1}

\begin{table*}
\begin{scriptsize}
  \begin{center}
 \begin{tabular}{ccccccccc}
  \toprule
      \toprule

     \multicolumn{1}{c}{}      &       
     \multicolumn{4}{c}{C$^{17}$O \j{2}{1}    }        &
     \multicolumn{4}{c}{C$^{17}$O \j{3}{2}     }  \\
     \cmidrule(r){2-5}
     \cmidrule(l){6-9}

      \multicolumn{1}{c}{WFS}      &
      \multicolumn{1}{c}{Peak $T_{mb}$}             &              
      \multicolumn{1}{c}{$v_{lsr}$}                &
      \multicolumn{1}{c}{$\Delta v$}                &
      \multicolumn{1}{c}{$\int T_{mb}dv$}     &
      \multicolumn{1}{c}{Peak $T_{mb}$}             &              
      \multicolumn{1}{c}{$v_{lsr}$} &
      \multicolumn{1}{c}{$\Delta v$}                &
      \multicolumn{1}{c}{$\int T_{mb}dv$}     \\

      \multicolumn{1}{c}{}  &
      \multicolumn{1}{c}{(K)}  &              
      \multicolumn{1}{c}{(km\,s$^{-1}$)}                &
      \multicolumn{1}{c}{(km\,s$^{-1}$)}                        &
      \multicolumn{1}{c}{(K\,km\,s$^{-1}$)}     &
      \multicolumn{1}{c}{(K)}  & 
      \multicolumn{1}{c}{(km\,s$^{-1}$)}                        &
      \multicolumn{1}{c}{(km\,s$^{-1}$)}                &
      \multicolumn{1}{c}{(K\,km\,s$^{-1}$)}   \\

      \midrule

1   &1.04  &-17.33    &2.14  &2.64 &1.80   & -17.42  & 2.06  &4.56 \\
3   &1.17  &0.10      &1.72  &3.01 &1.52   & 0.09    & 1.89  &3.78 \\
4   &0.57  &0.80      &2.40  &2.35 &$<$0.54    &-     &-   &- \\
6   &1.42  &5.76      &1.81  &3.48 &1.55   & 5.79    & 1.76  &3.51 \\
12  &1.46  &110.88    &2.47  &4.71 &-      &-        &-       &- \\
13  &0.72  &21.81     &3.36  &2.71 &-      &-        &-       &- \\
15  &1.65  &59.92     &2.86  &5.70 &-      &-        &-       &- \\
18  &1.55  &120.91    &3.01  &5.36 &-      &-        &-       &- \\
19  &1.83  &44.07     &2.70  &6.67 &-      &-        &-       &- \\
22  &*	   &*         &*     &*    &1.67   &84.65    &2.08    &3.92   \\
23  &2.80  &84.17     &2.12  &7.30 &-      &-        &-       &-     \\
24  &0.84  &76.48     &2.72  &2.72 &-      &-        &-       &-    \\
28  &2.75  &84.09     &2.13  &8.16 &-      &-        &-       &-     \\
29  &*	   &*         &*     &*    &2.61   &59.22    &1.67    &7.37 \\
33  &2.80  &110.09    &2.76  &8.81 &-      &-        &-       &-      \\
35  &0.81  &26.14     &3.77  &3.35 &-      &-        &-       &-      \\
37  &2.77  &105.46    &2.10  &7.39 &-      &-        &-       &-      \\
38  &0.91  &111.16    &1.55  &1.84 &-      &-        &-       &-      \\
39  &*     &*         &*     &*    &2.03   &96.37   & 2.08    &5.48  \\
41  &$<$0.44  &-      &-  &- &-      &-        &-       &-      \\ 
42  &0.77  &98.07     &2.90  &2.80 &-      &-        &-       &-      \\                
51  &*	   &*         &*     &*    &1.25   &84.52   &1.25     &2.49   \\ 
55  &0.91  &49.50     &3.96  &5.17 &-      &-        &-       &-      \\               
57  &0.71  &83.47     &1.99  &1.84 &-      &-        &-       &-      \\               
58  &1.58  &82.75     &4.25  &7.80 &-      &-        &-       &-       \\              
59  &1.22  &76.42     &2.56  &3.87 &-      &-        &-       &-       \\              
60  &$<$0.46  &-      &-  &- &-      &-        &-       &-       \\              
61  &1.99  &76.87     &2.72  &6.55 &-      &-        &-       &-       \\
62  &$<$0.36  &-     &-  &- &-      &-        &-       &-       \\              
63  &0.12  &50.64     &1.42  &0.86 &-      &-        &-       &-        \\             
64  &1.28  &10.40     &3.22  &4.75 &-      &-        &-       &-        \\              
66  &2.41  &66.13     &3.05  &9.04 &-      &-        &-       &-        \\             
67  &1.01  &32.62     &3.17  &3.96 &-      &-        &-       &-       \\              
68  &1.28  &55.09     &2.38  &4.14 &-      &-        &-       &-        \\             
69  &0.85  &54.54     &2.39  &2.20 &-      &-        &-       &-         \\            
70  &0.91  &6.54      &1.82  &2.55 &-      &-        &-       &-         \\            
71  &0.87  &14.17     &2.93  &3.14 &-      &-        &-       &-         \\            
72  &1.07  &3.77      &3.94  &4.86 &-      &-        &-       &-         \\            
74  &0.96  &4.82      &3.34  &3.77 &-      &-        &-       &-         \\            
75  &2.93  &23.09     &1.45  &5.97 &-      &-        &-       &-         \\            
76  &2.74  &24.04     &1.18  &4.75 &-      &-        &-       &-         \\ 
77  &2.81  &26.66     &1.05  &4.20 &5.03   & 26.72   & 0.89  &6.14        \\ 
78  &*     &*         &   *  &*    &1.12   & 21.49   & 2.32  &3.14        \\ 
79  &*	   &*         &   *  &*    &1.93   & 22.45   & 2.26  &5.86        \\ 
80  &1.55  &28.86     &1.85  &3.90 &2.10   & 29.01   & 1.86  &4.97         \\
81  &1.99  &20.29     &1.69  &4.38 &1.17   & 20.43   & 0.94  &2.63         \\
82  &3.13  &20.06     &1.81  &7.77 &4.23   & 19.96   & 1.48  &8.02         \\
83  &0.94  &21.47     &1.66  &2.43 &0.97   & 21.60    & 1.40   &1.95        \\
84  &1.19  &22.32     &2.01  &3.55 &1.42   & 22.54   & 1.49  &2.54         \\
86  &0.87  &5.72      &1.94  &2.55 &-      &-        &-       &-          \\        
89  &0.80  &5.47      &0.92  &1.04 &-      &-        &-       &-          \\       
90  &*	   &*         &  *   &*    &1.49   & -3.48   & 3.10    &6.52      \\ 
91  &*	   &*         &   *  &*    &4.96   & -1.89   & 1.10    &7.81      \\ 
92  &0.83  &-1.69     &1.63  &1.88 &-      &-        &-       &-         \\                    
93  &1.48  &-1.69     &1.60  &3.57 &-      &-        &-       &-           \\      
94  &1.36  &6.03      &2.61  &4.13 &-      &-        &-       &-           \\      
98  &1.83  &11.50     &1.03  &2.90 &-      &-        &-       &-           \\      
101 &1.81  &-18.38    &1.81  &4.41 &-      &-        &-       &-           \\      
102 &$<$0.65  &-      &-  &- &-      &-        &-       &-           \\      
103 &$<$0.69  &-    &-  &- &-      &-        &-       &-           \\      
104 &1.52  &-12.35    &0.95  &2.38 &-      &-        &-       &-           \\       
107 &*     &*         &*     &*    &1.83   & -45.97  & 1.57   &3.75          \\ 
108 &*     &*         &   *  &*    &1.51   & -53.21  & 2.70   &5.03         \\ 
109 &*     &*         &   *  &*    &1.49   & -44.37  & 2.58  &4.83          \\ 
110 &*     &*         &   *  &*    &1.80   & -54.75  & 2.00     &4.67        \\
111 &*	   &*         &   *  &*    &4.77   & -17.84  & 1.09  &7.00           \\
112 &*     &*         &   *  &*    &3.13   & -18.59  & 1.24  &4.95           \\

\hline                                                                                          
                                                    
\end{tabular}                                       
\end{center}                   
\end{scriptsize}
\caption{\footnotesize  Observed line parameters. The columns give the peak intensity,
velocity, linewidth and integrated intensity for \cseventeeno. The first three parameters are derived from hyperfine fitting  of the data  using CLASS. The integrated intensity was obtained using a  single Gaussian fit. Three sigma upper limits are shown for sources where a line was not detected above this threshold. An * indicates an data that has been been observed in the previous year and is given in Table \ref{tab:tablehfs}.}   
\label{tab:newcohfs}
\end{table*}

\linespread{1.4}

\subsection{Line Profiles}
\label{sec:fits}
The line parameters were initially measured using SPECX by fitting
Gaussian line-profiles to derive the source velocity ($v_{lsr}$), line
width ($\Delta v$), peak line flux ($T_A^*$) and integrated line
intensities.

All the C$^{18}$O data were fitted with a single Gaussian. These fits produced
typical peak temperatures of $T_A^* \sim 2.5-7.5$\,K, and linewidths of $\Delta
v\sim1.5-3.5$\,km\,s$^{-1}$.  However examination reveals that a significant
fraction of these C$^{18}$O data ($\sim$65\%) are distinctly better fit, 
  with smaller residuals, by the sum of two different velocity components.
The results of two components fits are also given in Table \ref{tab:lw}. Those
sources with multiple components can be divided into two categories; those
where the 2 components are of approximately equal width, but are offset in
velocity, and those where there is a definite separation into a broad and a
narrow component, with the broad component possibly being a related to the
  outflow from the source. In this latter case the typical linewidths of the
broad and narrow components are $\Delta v_{br}\ge3$\,km\,s$^{-1}$ and $\Delta
v_{n}\le 2$\,km\,s$^{-1}$.

\linespread{1}
\begin{figure}[!h]
\begin{center}
\includegraphics[angle=270,width=0.98\hsize]{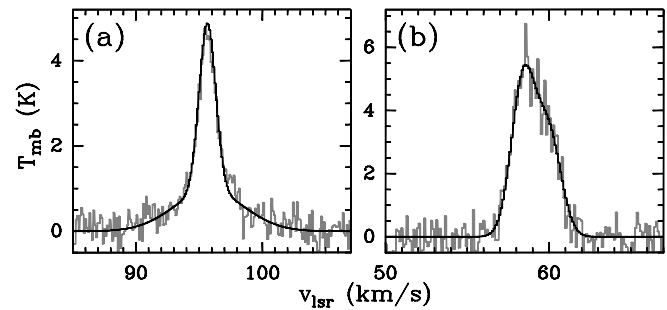}
\caption{\footnotesize Example spectra of \ceighteeno\ line profiles for
  {\bf(a)} WFS30 and {\bf(b)} WFS29 which are best fit by 2 Gaussian
  components shown as the smooth curve over each line.}
\label{fig:2comp}
\end{center}
\end{figure}
\linespread{1.4}

The integrated intensities for C$^{17}$O have been calculated from a Gaussian
fit using SPECX. However unlike the C$^{18}$O, the transitions of C$^{17}$O
have hyperfine structure. Simple Gaussian fits to the C$^{17}$O data therefore
overestimate the intrinsic velocity dispersion. The C$^{17}$O line parameters
were therefore fitted with the known hyperfine structure (hfs) of the
respective transitions using METHOD HFS in the CLASS package assuming all
components have equal excitation temperatures. For the \j{2}{1} transition
there are 9 hyperfine components, while there are 14 for the \j{3}{2}
transition.  The line parameters, the peak intensity, the $v_{lsr}$, the
linewidth and the integrated intensity are given in Tables \ref{tab:tablehfs}
and \ref{tab:newcohfs}. The optical depths calculated from the hfs fits imply
these sources are optically thin in C$^{17}$O with $\tau\sim0.1$.

\subsection{Ratio of Integrated Intensities}
\label{sub:intint}

The ratio of C$^{18}$O to C$^{17}$O line intensities can constrain the optical
depth of the line if the abundance ratio is known \citep*{ladd98}.  Penzias
conducted a survey of 15 massive star forming regions lying in the galactic
disk \citep{penzias81} and derived a value of R = [$^{18}$O]/[$^{17}$O] = 3.65
$\pm$ 0.15 and found no gradient based on galactocentric distance. Further
work by different authors have produced values, assuming that
[C$^{18}$O]/[C$^{17}$O] = [$^{18}$O]/[$^{17}$O], ranging from R = 2.9 $\pm$ 1.2
\citep{sheffer02} to 4.15 $\pm$ 0.52 \citep{bensch01}. More recently
\citet{ladd04} conducted a survey of 648 lines of sight towards 5 star forming
regions in the Taurus molecular cloud and combining the results for the
individual regions concluded that R = 4.0 $\pm$ 0.5, a value consistent with an
analysis of other clouds by \citet*{wouterloot05}.

\begin{figure}[!h]
\begin{center}
\includegraphics[angle=0,width=0.9\hsize]{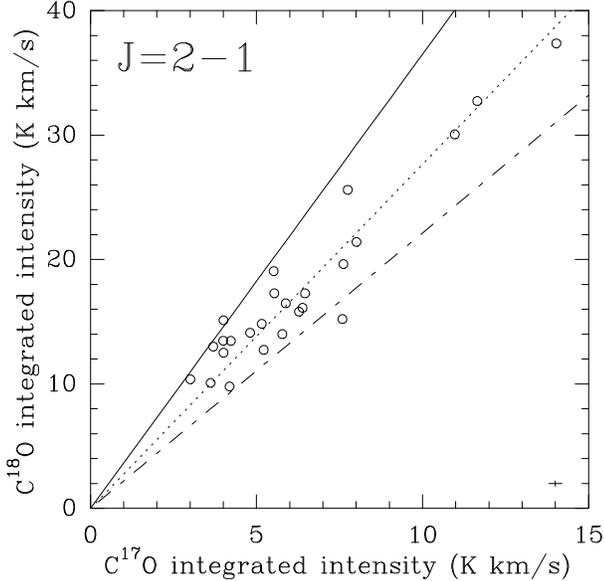}
\caption{{\footnotesize Comparison of the integrated intensities of \ceighteeno\
    and \cseventeeno. The solid line denotes an abundance ratio of
    \cseventeeno\ to \ceighteeno\ of 3.65, the expected value if both lines
    are optically thin \citep{penzias81,ladd98}. The dotted line indicates the
    best-fit curve of 2.8. The dashed line indicates an optical depth of 2
    associated with C$^{18}$O}}
\label{fig:ratio}
\end{center}
\end{figure}

A comparison of our measured \cseventeeno\ and \ceighteeno\ integrated
intensities is shown in Figure \ref{fig:ratio}.  Assuming equal excitation
temperatures and beam filling factors for the two species, if the emission is
optically thin one would expect the ratio of the integrated line intensity of
the two species to equal the abundance ratio.  A standard value of 3.65 for
the abundance ratio of \ceighteeno\ to \cseventeeno\ is indicated by the solid
line.  Of the sources, 5 lie approximately on the line implying that for the
\ceighteeno\ and \cseventeeno\ lines are optically thin.  The absence of any
sources above the line is consistent with both the abundance ratio values of
Penzias (3.65) and Ladd (4.0), but not higher values.  As so many of
our sources lie close to this line it suggests that the value of 3.65 is the
upper limit for this ratio.  However, the majority of our sources lie below
this value suggesting that the \ceighteeno\ emission is not optically thin. 
  The figure also shows the expected ratio of integrated intensities for
  optical depth 2 in the \ceighteeno\ transition. Due to the overlap of the
  hyperfine components of the \cseventeeno\  the optical depth at the line peak
  depends on the velocity dispersion. The line on the figure has been
  calculated for a FWHM velocity of 3\,km\,s$^{-1}$ (and a \ceighteeno\ to
  \cseventeeno\ abundance ratio of 3.65) giving a corresponding \cseventeeno\
  peak optical depth of 0.5. If the FWHM velocity was 1\,km\,s$^{-1}$, the
  \cseventeeno\ optical depth would be reduced by a factor of 0.67, whereas if
  the the FWHM velocity was 5\,km\,s$^{-1}$, the optical depth would be
  increased by a factor of 1.05.  The figure shows that the \cseventeeno\
  emission is not highly optically thick towards these sources, a result
  consistent with optical depth implied by the fit to the hyperfine structure
  of the observed lines (Sec.~\ref{sec:fits}).  The figure also shows that
  there is no correlation between line strength and optical depth.  

\linespread{1}
\linespread{1.4}

\subsection{Column Densities}

Since the \cseventeeno\ emission appears to not be optically thick, we
calculate the column densities assuming that the
\cseventeeno\ is optically thin. This is consistent with the analysis
of \cseventeeno\ towards similar objects by both \citet{vdt00} and
\citet{font06}.

To calculate the CO column densities we used the following general expression
assuming LTE and optically thin emission;
\begin{displaymath}
N_{tot}=\frac{3k}{8\pi^3\nu\mu^2g_u}\frac{Q(T_{ex})}{\exp(-E_u/kT_{ex})}\int
T_{mb}dv
\end{displaymath}
Here $g_u$ is the statistical weight, $Q(T)$ is the partition function
and $\mu$ is the electric dipole moment for that molecule.  Expressing
the integrated intensity in K\,km\,s$^{-1}$, the dipole moment in Debye
($\mu$=0.11\,D for CO) and the frequency in GHz, this reduces to
 
\begin{displaymath}
N_{tot}=1.67\times10^{14}\frac{Q(T_{ex})}{\mu^2({\rm Debye})\nu({\rm GHz})
  g_u}\exp\left(\frac{E_u}{T_{ex}}\right)\int T_{mb}dv 
\end{displaymath}
This method assumes a constant excitation temperature ($T_{ex}$) along
the line of sight which is a free parameter in the analysis.  For
  all their observed sources, SBSMW also estimated the temperature of
  the cold component of the dust; the mean over the whole sample was
  45\,K.  SBSMW also measured the gas temperature towards many of the
  sources observed here using NH$_3$. Over their entire sample SBSMW
  find a mean temperature of 19\,K and for all but two of the sources
  observed here they found temperatures of $<20$\,K.  To estimate the
  \cseventeeno\ column density we have adopted 30\,K, a compromise value
  intermediate between cold dust and NH$_3$ temperatures. The column
  density is of course sensitive to the assumed excitation
  temperature and is at a minimum at $T_{ex}$= 17\,K. At 10\,K and 30\,K
  the column densities are approximately equal and about 16\% higher
  than at the minimum. 

For comparison with the observations of the dust towards the sources it was
necessary to re-convolve the 850\,$\mu$m data to a beamsize to match that of the
CO beam. We first calculate the mass in the beam using the expression
\begin{displaymath}
M_{(beam)}=\frac{S_{\nu(beam)} d^2}{\kappa_\nu B_\nu(T_{dust})}
\end{displaymath}
where we take the value for the opacity from WFS04 as
$\kappa_{850}$=1.54$\times$10$^{-2}$\,cm$^2$g$^{-1}$ and the dust temperature
(T$_{dust}$) as the cold dust component given in SBSMW.  We then derive the
beam-averaged hydrogen column density from the submillimetre mass using the
relationship from \citet{hildebrand83} given as
\begin{displaymath}
M_{(beam)}=[\pi \theta^2_{\frac{1}{2}}d^2]N(H+H_2)m_H\mu
\end{displaymath}
where $m_H$ is the mass of a hydrogen atom and $\mu$ is the ratio of total gas
mass to hydrogen mass and $\theta_{\frac{1}{2}}$ is the beam radius. In both
steps the distance to the source, $d$, is taken to be the near-distance if
there is an ambiguity (SBSMW). The results for $N$(\cseventeeno) are given along
with the hydrogen column density derived from the dust continuum emission in
Table \ref{tab:newcoN}. 

\linespread{1}

\begin{table*}
\begin{scriptsize}
  \begin{center}
 \begin{tabular}{ccccccccccc}
  \toprule
      \toprule

     \multicolumn{1}{c}{}      & 
       \multicolumn{1}{c}{N(C$^{17}$O)}            &
     \multicolumn{1}{c}{N(C$^{17}$O)}            &
     \multicolumn{1}{c}{N(H$_2$)}            &
     \multicolumn{1}{c}{$\frac{\rm{N(C^{17}O}\ 2-1)}{\rm{N(H_2)}}$}            &
      \multicolumn{1}{c}{}      & 
        \multicolumn{1}{c}{N(C$^{17}$O)}            &
     \multicolumn{1}{c}{N(C$^{17}$O)}            &
     \multicolumn{1}{c}{N(H$_2$)}          &
     \multicolumn{1}{c}{$\frac{\rm{N(C^{17}O}\ 2-1)}{\rm{N(H_2)}}$}             \\
     \cmidrule(l){2-5}
     \cmidrule(l){7-10}

      \multicolumn{1}{c}{WFS}      &
       \multicolumn{1}{c}{\j{2}{1}}                &
      \multicolumn{1}{c}{\j{3}{2}}                &
      \multicolumn{1}{c}{}         &
      \multicolumn{1}{c}{} &
      \multicolumn{1}{c}{WFS}      &
       \multicolumn{1}{c}{\j{2}{1}}                &
      \multicolumn{1}{c}{\j{3}{2}}                &
      \multicolumn{1}{c}{} &
      \multicolumn{1}{c}{}  \\

      \multicolumn{1}{c}{}      &
         \multicolumn{1}{c}{($\times$10$^{15}$)}                &
      \multicolumn{1}{c}{($\times$10$^{15}$)}                &
      \multicolumn{1}{c}{($\times$10$^{22}$)}         &
      \multicolumn{1}{c}{($\times$10$^{-8}$)}         &
      \multicolumn{1}{c}{}      &
         \multicolumn{1}{c}{($\times$10$^{15}$)}                &
      \multicolumn{1}{c}{($\times$10$^{15}$)}                &
      \multicolumn{1}{c}{($\times$10$^{22}$)}  &
      \multicolumn{1}{c}{($\times$10$^{-8}$)}           \\

      \midrule
                               
1     &0.64   &0.90   &4.03  &1.58     &68   &1.00  &-     &2.86   &3.50 \\    
3     &0.73   &0.74   &2.87  &2.53     &69   &0.53  &-     &-      &-    \\    
4     &0.57   &0.00   &2.38  &2.38     &70   &0.62  &-     &1.90   &3.24  \\         
6     &0.84   &0.69   &3.57  &2.35     &71   &0.76  &-     &3.34   &2.27  \\       
12    &1.14   &-      &4.16  &2.73     &72   &1.21  &-     &9.64   &1.25   \\       
13    &0.65   &-      &10.01 &0.65     &74   &0.91  &-     &6.97   &1.31  \\         
14    &1.87   &-      &8.15  &2.30     &75   &1.44  &-     &5.96   &2.42   \\      
15    &1.37   &-      &1.62  &8.47     &76   &1.15  &-     &2.03   &5.66\\           
16    &3.39   &-      &12.09 &2.80     &77   &1.01  &1.21  &2.11   &4.81  \\  
17    &2.81   &-      &4.96  &5.67     &78   &0.98  &0.62  &3.60   &2.72 \\   
18    &1.29   &-      &6.09  &2.13     &79   &1.33  &1.15  &12.15  &1.10 \\   
19    &1.61   &-      &23.25 &0.69     &80   &0.94  &0.98  &4.09   &2.30\\    
20    &1.01   &-      &0.98  &10.28    &81   &1.06  &0.52  &1.80   &5.87 \\   
21    &1.26   &-      &1.20  &10.47    &82   &1.88  &1.58  &3.06   &6.13 \\      
22    &1.93   &0.77   &5.53  &3.50     &83   &0.59  &0.38  &1.54   &3.81   \\   
23    &1.76   &-      &-     &-        &84   &0.86  &0.50  &1.62   &5.29 \\   
24    &0.66   &-      &2.50  &2.63     &85   &1.07  &-     &2.53   &4.25  \\  
25    &1.84   &-      &7.81  &2.36     &86   &0.54  &-     &1.00   &5.43  \\   
28    &1.97   &-      &4.31  &4.57     &87   &1.22  &-     &2.73   &4.47  \\       
29    &1.66   &1.45   &3.06  &5.99     &88   &1.64  &-     &2.73   &6.01  \\       
30    &0.89   &-      &4.44  &2.01     &89   &0.25  &-     &0.91   &2.77 \\   
33    &2.13   &-      &6.13  &3.47     &90   &1.34  &1.29  &8.40   &1.59\\    
34    &2.65   &-      &4.31  &6.15     &91   &1.40  &1.54  &2.53   &5.52  \\  
35    &0.81   &-      &6.13  &1.32     &92   &0.50  &-     &1.34   &3.74   \\ 
36    &1.54   &-      &3.24  &4.76     &93   &0.86  &-     &2.91   &2.96   \\   
37    &1.78   &-      &3.47  &5.15     &94   &1.00  &-     &6.77   &1.47  \\    
38    &0.43   &-      &2.33  &1.85     &95   &0.68  &-     &1.15   &5.88  \\       
39    &1.56   &1.08   &3.63  &4.30     &96   &1.23  &-     &3.60   &3.42  \\    
41    &-   &-      &0.08  &-           &97   &1.06  &-     &2.86   &3.71  \\      
42    &0.68   &-      &0.87  &7.74     &98   &0.70  &-     &-      &- \\         
51    &1.25   &0.49   &1.36  &9.17     &99   &1.02  &-     &5.28   &1.93  \\  
55    &1.25   &-      &3.28  &3.18     &100  &1.16  &-     &-      &-     \\       
57    &0.44   &-      &2.66  &1.67     &101  &1.06  &-     &3.29   &3.24 \\   
58    &1.88   &-      &7.69  &2.45     &102  &-  &-     &-      &-   \\    
59    &0.93   &-      &3.86  &2.42     &103  &-  &-     &2.31   &-    \\        
60    &-   &-         &0.78  &-        &104  &0.57  &-     &1.38   &4.15   \\        
61    &1.58   &-      &5.82  &2.72     &107  &0.73  &0.74  &2.88   &2.53  \\        
62    &0.24   &-      &1.11  &2.18     &108  &1.43  &0.99  &6.66   &2.13    \\
63    &0.21   &-      &0.47  &4.35     &109  &0.97  &0.95  &7.75   &1.25    \\
64    &1.15   &-      &4.05  &2.83     &110  &0.87  &0.92  &2.69   &3.26    \\
66    &2.18   &-      &2.86  &3.83     &111  &1.51  &1.38  &2.67   &5.68    \\
67    &0.96   &-      &6.31  &1.51     &112  &0.97  &0.98  &2.67   &3.64    \\

\hline
                                                    
\end{tabular}                                       
\end{center}                   
\end{scriptsize} 
\caption{\footnotesize Column densities calculated from the line
  intensities with an average rms of $\sim$0.2\,K. The hydrogen column density
  corresponds to a 20$''$ beam to directly compare with the \j{2}{1} transitions
  of each isotope.}  
\label{tab:newcoN}
\end{table*}


\linespread{1.4}

\begin{figure}
\begin{center}
\includegraphics[angle=0,width=0.9\hsize]{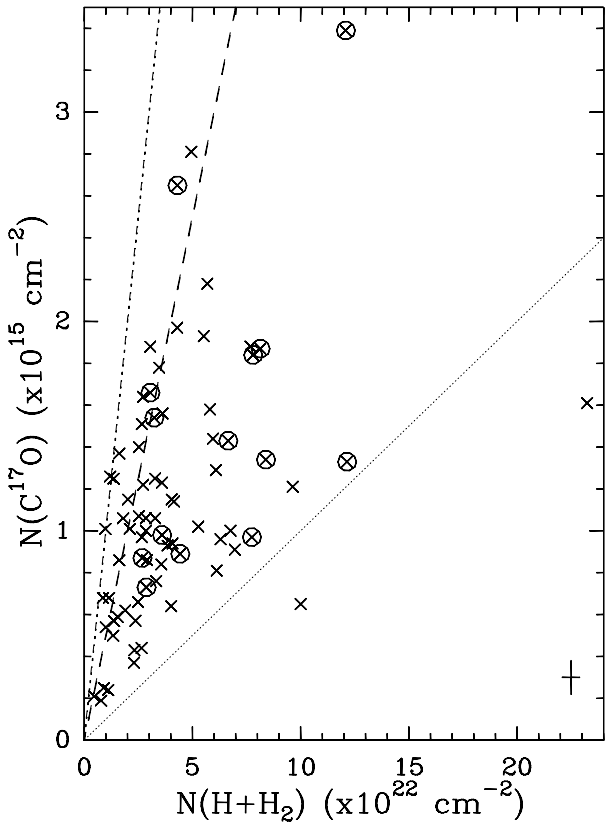}
\caption{\footnotesize Comparison of the column density C$^{17}$O with the
  H$_2$+H column density derived from the dust continuum emission. Typical
  uncertainties are shown in the lower right corner of the panel. The dashed
  lines represent lines of constant abundance: Short dashed =
  5$\times$10$^{-7}$, dot-dashed = 1$\times$10$^{-7}$, long dashed =
  5$\times$10$^{-8}$ \& dotted = 1$\times$10$^{-8}$. The circled sources
  indicate those which have been modelled (see Sec.\ref{sub:modelling}).}
\label{fig:Nfigs}
\end{center}
\end{figure}

Figure \ref{fig:Nfigs} shows the column density C$^{17}$O plotted against that
calculated for H$_2$ from the dust continuum emission.  Also shown are lines
of constant CO abundance chosen to constrain the data. It is immediately
evident that these graphs show a large scatter in the abundances. The
C$^{17}$O data are constrained by abundances of $\sim$1$\times$10$^{-7}$ and
$\sim$7$\times$10$^{-9}$, a factor of approximately fourteen spread in the
abundance.  Taking the standard [C$^{17}$O]/[H$_2$] abundance to be 4.7$\times
10^{-8}$ \citep{frerking82}, we would expect our sources to lie either on this
line or below it in the case of CO depletion, yet it is clear  that
this value does not represent the upper limit for our sources and a
significant fraction lie up to a factor of 2 above this line.

  To explore the reliability of the derived \cseventeeno\
  abundances, we have also run detailed models of the emission
  expected from these sources.  These models allow us to investigate
  the possible consequences of some of the assumptions made when
  calculating the column densities, particularly the assumed
  excitation temperature and dust temperature, together with the role
  of density gradients within the beam.


\linespread{1}
\linespread{1.4}

\section{Modelling}
\label{sub:modelling}

To model the emission of \cseventeeno\ and \ceighteeno\ we used the 1D
radiative transfer code RATRAN developed by \citet{ratran}. RATRAN
utilises the Monte Carlo method to calculate the radiative transfer of
molecular lines through a dusty shell.

\citet*[hereafter WFS05]{wfs05} modelled the 850\,$\mu$m emission for a
selection of the sources observed by WFS04 using the 1D
radiative transfer code DUSTY \citep*{dusty} in order to determine the
physical parameters of the envelopes around the central embedded
sources.  Of those sources for which we have both \cseventeeno\ and
\ceighteeno\ data, 14 were successfully modelled by WFS05. We have
utilised the parameters of the best-fit models from WFS05 to generate
the input for RATRAN.

WFS05 identified certain sources as asymmetric and used radial slices
to explore the structure in various directions, as a result four of
our 14 sources actually have multiple models in WFS05.  For these
sources we correspondingly generated multiple models for RATRAN. These
are indicated by alphabetical suffixes to the source names.

As RATRAN is a 1D code it models the sources as spherically symmetric and
hence neglects any openings or elongations due to outflows or winds. The model
assumes a dust-free cavity surrounding the star from radius $r=0$ to
$r=r_{1}$. Between $r_{1}$ and $r_{2}$ lies the envelope to be modelled. We
divided the radial distance between r$_{1}$ and r$_{2}$ into 30
logarithmically spaced shells to give increased resolution towards the inner
region of the shell where the temperature and density profiles change most
rapidly. Thirty shells were used corresponding to the number used by
\citet[hereafter VVEB]{vdt00} in their RATRAN modelling of massive young
stars.  We also ran a number of identical models using both 15 and 60 shells
and found a negligible effect on the output compared to the 30 shell models.

To determine the input for RATRAN we used the following parameters
derived in WFS05: power-law index ($\alpha$) of the envelope density
profile $n(r$) where $n\propto r^{-\alpha}$, the temperature at the
inner boundary $T_1$, the scale of the envelope defined by the inner
radius, $r_{1}$ and $Y$, where $Y$ is the ratio between $r_2$ and $r_1$. The outer radius can then be calculated from $r_2 = Yr_1$.

\linespread{1} 
\begin{table}
\begin{scriptsize}
  \begin{center}
 \begin{tabular}{@{}ccccccc@{}}
  \toprule
    \toprule     
 
     \multicolumn{1}{c}{WFS}      &   
     \multicolumn{1}{c}{$r_{1}$}      & 
     \multicolumn{1}{c}{$r_{2}$}      &
     \multicolumn{1}{c}{$\alpha$}      & 
     \multicolumn{1}{c}{$n(r_1)$}      &    
     \multicolumn{1}{c}{$\Delta v$(C$^{17}$O)}      & 
     \multicolumn{1}{c}{$\Delta v$(C$^{18}$O)}      \\
     \cmidrule(l){2-3}
     \cmidrule(l){4-7}

    \multicolumn{1}{c}{}      &    
    \multicolumn{1}{c}{10$^{16}$cm}      & 
    \multicolumn{1}{c}{10$^{18}$cm}      &   
    \multicolumn{1}{c}{}      &
    \multicolumn{1}{c}{10$^6$cm$^{-3}$}      & 
    \multicolumn{1}{c}{km\,s$^{-1}$}      &    
    \multicolumn{1}{c}{km\,s$^{-1}$}      \\  
   
   \midrule
14    &3.02  &1.51   &1.5 &3.09  & 2.84 &2.21 \\
16    &4.06  &4.06   &2.0 &8.74   & 3.35 &2.95 \\
25    &2.69  &5.39   &1.5 &1.33  & 2.64 &2.20 \\
29    &2.27  &1.59   &1.0 &0.48  & 2.59 &2.17 \\
30    &0.23  &1.84   &1.0  &4.41  & 2.37 &2.27 \\
34a   &1.84  &0.18   &1.0  &32.97  & 3.71 &3.30  \\
34b   &0.63  &5.04   &1.5  &2.19  & 3.71 &3.30 \\  
36    &0.45  &0.90   &1.0   &0.83  & 2.94 &2.32 \\  
79    &2.41  &1.21   &1.5 &6.78  & 2.71 &2.33 \\  
90a   &1.78  &1.25   &1.5 &3.44  & 3.27 &2.87 \\  
90b   &1.83  &1.83   &2.0 &12.74  & 3.27 &2.87 \\  
107   &3.50  &2.45   &0.5  &0.05  & 2.24 &1.59  \\ 
108a  &0.64  &2.56   &1.5  &8.35  & 3.25 &2.83 \\  
108b  &0.22  &3.52   &1.0  &1.07  & 3.25 &2.83 \\  
109   &0.23  &3.68   &1.5  &33.80   & 2.54 &2.00 \\  
110a  &0.64  &2.56   &1.5  &5.19  & 2.82 &2.09 \\  
110b  &1.70  &3.40   &1.5  &0.80  & 2.82 &2.09 \\  
110c  &1.83  &1.83   &2.0 &16.68  & 2.82 &2.09 \\

\hline
                                                    
\end{tabular}                                       
\end{center}                 
\end{scriptsize} 
\caption{\footnotesize Parameters used in the detailed modelling of the
  sources. Where $r_{1}$ is the inner radius of the modelled envelope, $r_{2}$ is the outer radius and $\alpha$ is the power-law index of the number
  density profile $n(r)$ where $n\propto r^{-\alpha}$. Column 5 gives the hydrogen number density at $r_{1}$. } 
\label{tab:data}                   
\end{table}

 \linespread{1.4} 

\begin{figure*}
\includegraphics[angle=270,width=0.95\hsize]{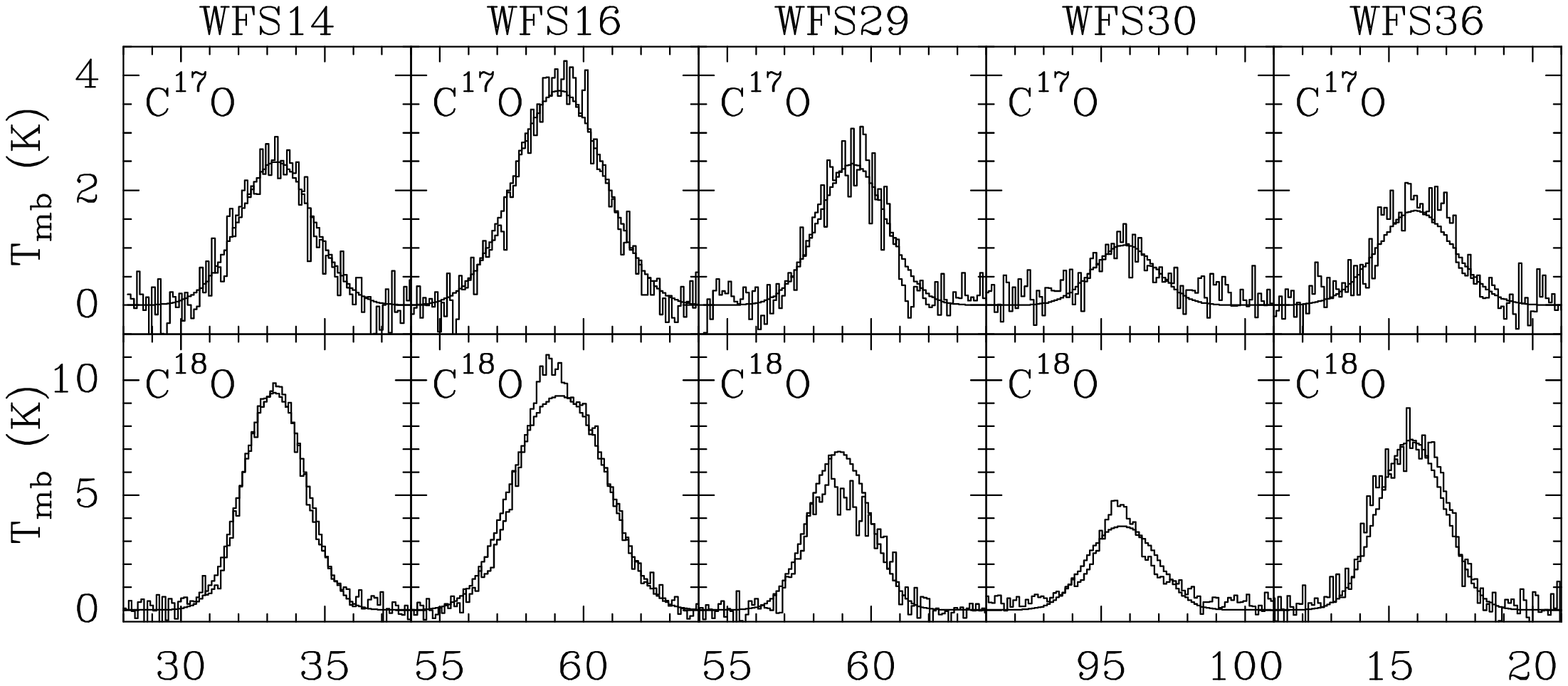}%
\vspace{0.1in}
\includegraphics[angle=270,width=0.77\hsize]{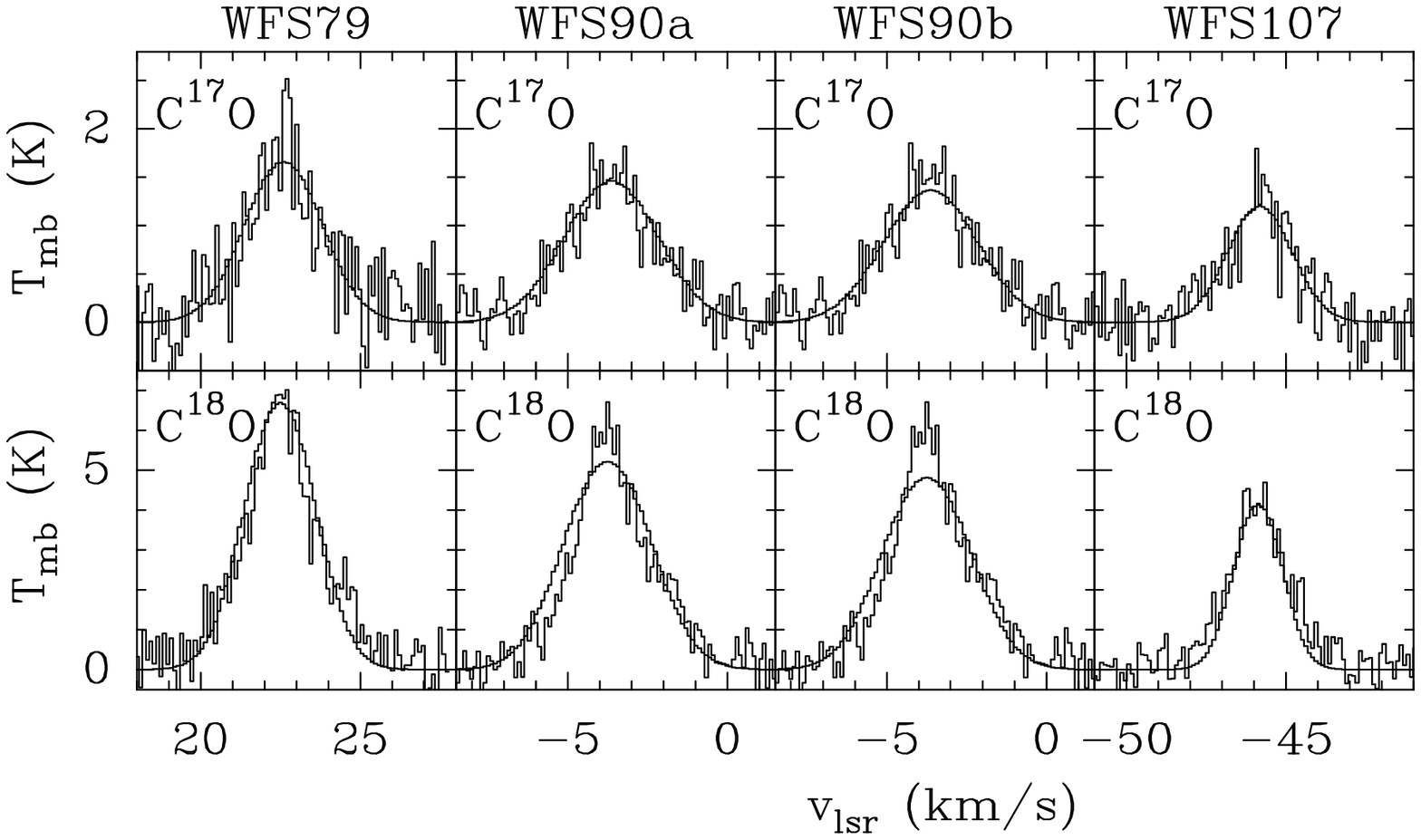}\caption{\footnotesize  Comparison of the \cseventeeno\ \j{2}{1} (upper spectra) and \ceighteeno\  \j{2}{1} (lower spectra) spectra with the results of the best fit results of the RATRAN models (solid curves). The model parameters are given in Tables  \ref{tab:data} and \ref{tab:bf}.}
\label{fig:overlay1}
\end{figure*}

We generated the density profile for each model by combining the
power-law index ($\alpha$) of the envelope density profile given in
WFS05 with the density at $r_{1}$ which we calculate from the
850\,$\mu$m mass and volume of the envelope (WFS04). As the dust temperature
profile is not given in WFS05 we regenerated the WFS05 models using
DUSTY to determine the individual temperature profile for each model.
This profile was then interpolated to the radii of the 30 shells used
to trace the envelope. Given the high densities in these regions we
have assumed that the gas kinetic temperatures and the dust
temperatures are equal.

The turbulent line widths were taken from Gaussian fits of the
optically thin C$^{17}$O data. The presence of hyperfine structure for
the C$^{17}$O was neglected for the models.  By fixing the linewidth
from our observations we were able to isolate the CO abundance
relative to hydrogen as the free parameter.  For simplicity we
restricted our models to a simple uniform abundance of C$^{17}$O
throughout the envelope of each source.  The input parameters for each
model are given in Table \ref{tab:data}.

These parameters were then used in conjunction with the adopted
collisional rate coefficients of \citet{flower01} \citep[see
also][]{schoier05} to generate the predicted line emission. The \cseventeeno\
abundances were adjusted to produce the best match to the observed
\cseventeeno\ line profiles. The modelled sources are circled in Figure
\ref{fig:Nfigs}, illustrating that they cover a range
dust and gas column densities and a range of implied \cseventeeno\
abundances.

The output spectra from RATRAN were continuum subtracted and convolved with
a Gaussian with a FWHM equal to the appropriate JCMT beam for
comparison with the observations.  We analysed the quality of the
models by using the reduced $\chi^2$ method, aligning the central
channel of the model spectra with the peak of the Gaussian fit for the
data. Figure~\ref{fig:overlay1} compares the observations with
the results of the best fit models.

\subsection{Modelling Results}
\label{sec:models}

\begin{figure}
\begin{center}
\includegraphics[angle=270,width=0.9\hsize]{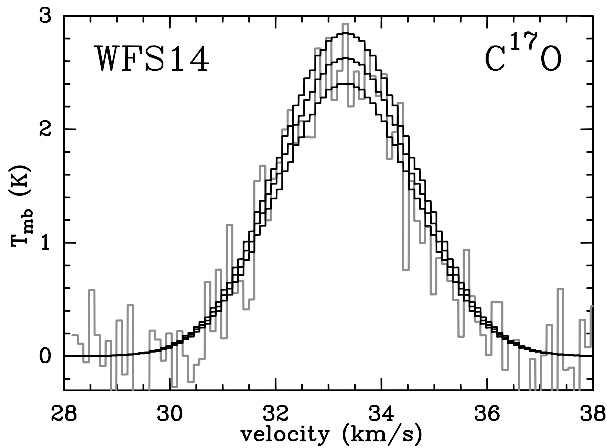}
\caption{\footnotesize The observed \cseventeeno\ \j{2}{1} emission
    towards WFS14 overlaid with the results of three models. The only
    difference between the three models is the assumed \cseventeeno\
    abundances of 5$\times 10^{-8}$, 5.5$\times 10^{-8}$ \& 6$\times 10^{-8}$. The abundance is constrained within the noise level of the observations to within $\sim$10\%.}
\label{fig:varychi}
\end{center}
\end{figure}

Figure \ref{fig:varychi} shows the success with which these (simple) models
can match the observed \cseventeeno\ line profiles and also the sensitivity of
the fit to the assumed \cseventeeno\ abundance. In this case the \cseventeeno\
abundance is constrained within the noise level of the observations to
within $\sim10$\%.

Since RATRAN calculates the continuum emission in parallel with the line
emission it was possible to check on the consistency of the models by
comparing the predicted 850\,$\mu$m flux with that measured.  This is especially
important in this work where we are looking at the ratio of the line to
continuum emission. For a number of sources modelled the predicted and
measured 850\,$\mu$m flux differed by more than a factor of 3. This is not
completely surprising as the best fit models of WFS05 were selected to provide
the best overall fit to the source SED and 850\,$\mu$m distribution, as such
they do necessarily provide excellent fits to the 850\,$\mu$m flux alone.
Therefore we have only considered the 8 sources where the predicted 850\,$\mu$m
continuum emission differs from the measured value by less than a factor of 3.
The results of these sources are given in Table \ref{tab:bf}.

\subsubsection{C$^{17}$O Analysis}

A comparison between the observed line profiles and the best fit model
results for all the modelled sources is shown in
Figure~\ref{fig:overlay1}.  The models can successfully match the
observed C$^{17}$O line profile with a reasonable range of C$^{17}$O
fractional abundances ranging over the sample from
1.8$\times$10$^{-8}$ to 1.5$\times$10$^{-7}$. Importantly for the
analysis of the sources for which models of the continuum emission do
not exist, comparison of the C$^{17}$O abundance inferred by the
modelling and that derived directly from the observations agree within
a factor of $\sim$2 (Table \ref{tab:bf}).  This result indicates that
the assumptions made in directly estimating the \cseventeeno\
abundance from the observations are not overly biasing the results.

\linespread{1}
\linespread{1.4}

\subsubsection{C$^{18}$O Analysis}

For completeness we also attempted to model the C$^{18}$O lines for the
sources modelled in C$^{17}$O.  Initial modelling of the C$^{18}$O lines
involved taking the best-fit C$^{17}$O abundance and scaling it up by a range
of potential values of the \ceighteeno\ to \cseventeeno\ abundance ratio; these were R$=$2.8, 3.65, 4.0 \& 4.5  (as mentioned in Sec. \ref{sub:intint}).  All
other free parameters remain constant with the exception of the velocity
dispersion which was adopted from the C$^{17}$O hfs fit (as the best estimate
of the intrinsic velocity dispersion within these regions). The results are
also shown in Figure \ref{fig:overlay1}. On inspection it is clear that a
number of the models suffer from high optical depths, resulting in a flattening
and broadening of the line shape, which is not seen in the observations.  The
inconsistencies between the data and the models for this isotopologue is
perhaps not unexpected given the two velocity component profiles towards many
of the sources which suggest that more than a simple envelope is contributing
to the line profile.

An example of such a scenario would be a central core (as
represented in our models), but surrounded by an extended low density
envelope with the \ceighteeno\ having different optical depths in the
envelope and in the core. The success of modelling the \cseventeeno\
from the core (Sec.~\ref{sec:models}) would suggest that the core has
the higher optical depth.

For sources with two component C$^{18}$O line profiles (Table
\ref{tab:lw}), the C$^{17}$O line is often intermediate in width
between the two C$^{18}$O components. This could also point towards
the presence of two different components of material. If there are
indeed two components then the simple column density analysis and
model are actually overestimating the actual C$^{17}$O and
\ceighteeno\ column densities in the core,  as some of the column
  density attributed to the core in a single component model is
  actually associated with the second component.   This suggests that
the abundances of these species in the core could be even lower than
the values derived here indicate.

It is possible to test this prediction with  maps of these
sources; C$^{18}$O emission extending beyond the detected 850\,$\mu$m
emission would provide evidence for an outer envelope. It is possible
that the relatively small chopper throw of the 850\,$\mu$m observations
may have artificially removed this component from the SCUBA
observations. For some sources this resulting underestimation of dust
emission could account for the high abundances seen towards a few
sources.

\linespread{1}
\begin{table}
\begin{small}
\begin{center}
 \begin{tabular}{@{}cccccc@{}}
    \toprule
    \toprule     
 
     \multicolumn{1}{c}{WFS}      &   
     \multicolumn{1}{c}{f(C$^{17}$O)}      &
     \multicolumn{1}{c}{$\chi^2_{red}$}      & 
     \multicolumn{1}{c}{[$^{18}$O]/[$^{17}$O]}      &
     \multicolumn{1}{c}{R(H$_2$)}  &
     \multicolumn{1}{c}{R(C$^{17}$O)}  \\
     &   
     ($10^{-8}$)      \\
   \midrule
14    &5.2  &1.69  &4.0  &1.36 & 1.35 \\
16    &10.1 &1.52  &4.0  &1.70 & 2.16 \\
29    &10.0 &1.79  &2.8  &0.76 & 1.00 \\
30    &5.0  &1.09  &4.5  &2.86 & 1.49 \\
36    &15.0 &1.58   &4.5 &1.30 &  1.88 \\  
79    &1.8  &2.38   &3.65 &0.98 &  0.98 \\  
90a  &4.5  &0.84  &3.65 &2.64 & 1.69 \\
90b   &4.5  &0.86  &3.65 &2.15 & 1.69 \\  
107   &4.0  &1.13  &2.8  &1.04 & 0.94 \\ 

\hline
                                                    
\end{tabular}                                       
\end{center}                 
\end{small} 
\caption{\footnotesize {
    Parameters of the best fit RATRAN models.  
    Column 2 is the fractional abundance of C$^{17}$O required for the
    model to match the observed \j{2}{1} line profile. Column 3 is the
    reduced $\chi^2$ between the observed \cseventeeno\ \j{2}{1}
    transition and the  modelled emission.  
    R(H$_2$) is the ratio of measured 850\,$\mu$m flux to that 
    generated by RATRAN. R(C$^{17}$O) is the ratio of the best fit abundance
    from the modelling to the abundance derived from comparing the data to the
    850\,$\mu$m hydrogen abundance:
    [C$^{17}$O/H$_2$]$_{mod}$/[C$^{17}$O/H$_2$]$_{data}$. Column 4 presents
    the best fit ratio of  \ceighteeno/\cseventeeno\  needed to 
     match the \ceighteeno\ observations, scaling up the best-fit
    \cseventeeno\ abundances listed. 
    The model line profiles are compared with the observations in 
    Figure \ref{fig:overlay1}.}}
\label{tab:bf}                   
\end{table}

 \linespread{1.4}

\section{Discussion}
\label{sec:discussion}

  The determination of the \cseventeeno\ column densities and hence
  abundances contain a number of sources of uncertainty. Since both
  the \cseventeeno\ lines are detected with good signal to noise
  ratios (Table \ref{tab:tablehfs}; Figure \ref{fig:overlay1}), and the
  dust continuum fluxes of the sources are well measured (WFS04),
  the statistical uncertainties due to noise are small.   The
    calibration uncertainties may be larger, but are likely to affect all
    the sources equally.  However the results do also depend on a
  number of other assumptions.  Ideally all the sources observed
should be modelled in detail to determine their \cseventeeno\
abundance throughout their envelopes and cores. However temperature
and density profiles are not (yet) available for all the objects, so
this is not currently practical.   Nevertheless the objects which
  have been modelled in detail do suggest that systematic issues do
  not dominate the results.  The similar \cseventeeno\ abundances
derived from the modelling of a subsample of the sources which span a
range of source properties and the \cseventeeno\ abundances derived
directly from the observations, suggests that despite the
approximations and assumptions it requires, the direct derivation
produces reasonable estimates of the average \cseventeeno\ abundance
towards the sources. It is these abundances on which we will focus.

To determine the origin of the scatter seen in Figure \ref{fig:Nfigs} we must
investigate the possible causes ranging from underlying physical differences
to the simplifying assumptions.  The consistency within a factor of 2 of the
 C$^{17}$O abundances generated using RATRAN and those derived straight from the data
suggests that the scatter between the observed C$^{17}$O column density and
H$_2$ column density is not a result of an overly simplistic data analysis or
the assumed excitation temperature.

It is important to recognise that the hydrogen column density derived
from the dust emission suffers from uncertainties due to the assumed
dust properties. In calculating the dust column density we have
adopted a constant value of $\kappa$, the dust mass opacity, for all
the sources. For our calculations we adopted a value from
\citet{ossenkopf} assuming a gas density of 10$^6$ cm$^{-3}$, with
thin ice mantles and 10$^5$ years of coagulation.  However we
investigated the influence of this assumption by recalculating
N(H$_2$) for the full range of possible opacities from Ossenkopf \&
Henning. By manipulating the opacity it is possible to move the entire
sample in Figure \ref{fig:Nfigs} to a range either above the
canonical C$^{17}$O abundance of 4.7$\times$10$^{-8}$ or below it.
However assuming that {\em all} the sources should lie at this value,
it is clear that changing the dust in order to allow that to happen
would have to be done on a source by source basis and utilising the
full range of potential opacities.

For those sources that lie to the left of the canonical
abundance line, an opacity corresponding to grains that have not undergone any
coagulation would need to be adopted. This lack of grain coagulation would
seem to imply that these sources, and the dust surrounding them, are young,
not yet having had time for the dust to coagulate.  To the other side of the
line lie the sources which appear to be the most depleted of our sample. To
change their abundances would require an opacity corresponding to some degree
of coagulation and for the most under-abundant sources this would need to
include an absence of ice mantles. The scenario of coagulated grains without
ice would be consistent with the dust close to a star which has heated its
surroundings and evaporated all icy mantles. However the temperature profiles
derived by DUSTY modelling show that only a small part of the envelopes are this warm, with the bulk of each envelope lying at temperatures below 50\,K.  Under these
conditions we would consider the complete absence of ice mantles to be an
unreasonable assumption.  The dust models indicate if mantles of any kind are
present, they would limit the change in the inferred hydrogen column density
to not more than 12\% of the calculated values.

In addition, \citet{vdt99} found that by comparing the dust mass from FIR and
submm data to the gas mass derived from regions without depletion of
C$^{17}$O, only the opacities for dust grains \emph{with} ice mantles
reproduced the standard dust-to-gas ratio of 1:100.  Additionally figure
\ref{fig:Nfigs} shows a number of the sources having abundances in
excess of 4.7$\times 10^{-8}$, which if not due to particular
circumstances for these objects, could indicate that for the sample as
a whole the hydrogen column density could be underestimated. Given
these points we conclude that the scatter seen between the abundances
does indeed reflect physical differences between the sources.

 In principle selective photodestruction of the \cseventeeno\ in the inner
   regions of the sources close to the central heating sources could give rise
   to the varying \cseventeeno\ abundance. However selective photodestruction
   can only be an issue in the region up to optical depth $\sim1$ in the UV
   from the inner edge of the circumstellar envelope.  Although this optical
   depth can be contributed by either dust or line self-shielding, since the
   circumstellar envelopes are dusty, it is the dust that will be the dominant
   UV opacity source.  This UV opacity corresponds to a visual optical depth
   $\sim0.1$ in the optical.  Since this corresponds to about 1\% of the
   thickness of the envelope (WFS04), the internal UV flux can not be
   affecting the \cseventeeno\ abundance by more than $\sim1$\%.
 
  The low optical depth of the \cseventeeno\ transitions suggest that if
   \cseventeeno\ is uniformly abundant throughout each source, the
   \cseventeeno\ emission, and hence \cseventeeno\ column density should, like
   the dust emission, trace the total column density even though the
   \cseventeeno\ has a relatively low critical density.

  The most likely explanation for the variation in the
   \cseventeeno\ abundance appears to be that it reflects different
   gas phase abundance of \cseventeeno\ due to its freeze out onto
   dust grains.  Such depletion of \cseventeeno\ (and by implication,
   CO and possibly other species) towards similar high mass sources
   has been previously reported by VVEB and \citet{font06}. 
 
{ \linespread{1} 
\begin{table}
\begin{scriptsize}
  \begin{center}
 \begin{tabular}{cccccccc}
  \toprule
    \toprule     
 
     \multicolumn{1}{c}{WFS}      &   
     \multicolumn{1}{c}{$\overline{T}_{mw}$}       &
 \multicolumn{1}{c}{WFS}      &   
     \multicolumn{1}{c}{$\overline{T}_{mw}$}       &
 \multicolumn{1}{c}{WFS}      &   
     \multicolumn{1}{c}{$\overline{T}_{mw}$}       &
    \multicolumn{1}{c}{WFS}      &   
     \multicolumn{1}{c}{$\overline{T}_{mw}$}       \\

   \midrule
6   & 23.2  & 19  & 23.5 & 66 & 26.5    &95   &29.7\\		
13  & 25.4  & 25  & 18.6 & 78 & 18.6   &107  &35.3\\			
14  & 28.5  & 29 & 34.1	 & 79 & 28.4    &108  &20.8\\	
16  & 23.4  & 30 & 23.4	 & 80 & 27.9    &109  &18.4\\		
18  & 18.6  & 36 & 30.0	 &90   & 25.4   &110  &20.8\\     

\hline
                                                    
\end{tabular}                                       
\end{center}                 
\end{scriptsize} 
\caption{\footnotesize Mass-weighted temperatures in Kelvin, calculated for those sources for which the data are available.}
\label{tab:mwt}                   
\end{table}

 \linespread{1.4}}

\subsection{Depletion of C$^{17}$O}

If the scatter in the C$^{17}$O abundance is indeed due to depletion, then one
might expect that the regions with the coolest dust would have the highest
depletion. VVEB investigated this using the mass-weighted temperature,
$\overline{T}_{mw}$, to define a characteristic temperature for each source.

\begin{figure}
\begin{center}
\includegraphics[angle=0,width=0.85\hsize]{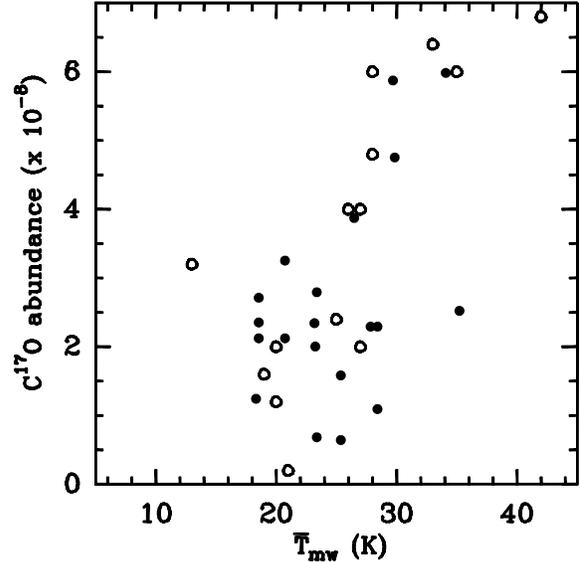}
\caption{{\footnotesize  C$^{17}$O (\j{2}{1}) abundance against the mass-weighted
    temperature, $\overline{T}_{mw}$. The solid dots denote our data whilst
    the circles represent the data from Figure 11 in VVEB.}}    
\label{fig:mwt}
\end{center}
\end{figure}

Figure \ref{fig:mwt} shows the C$^{17}$O abundance against the mass-weighted
temperature for all the sources for which the necessary parameters were
available. The mass-weighted temperature is defined by VVEB and was calculated
here using the temperature profile from the DUSTY modelling of the sources.
The figure shows data points from both our sources and those from VVEB. The
two data sets are in reasonable agreement, with the distribution of both being
consistent with depletion at low temperatures.  The Spearman correlation
coefficient indicates that the \cseventeeno\ abundance and mass-weighted
temperature are correlated at a 96\% confidence level.

We do however see a number of our sources exhibiting higher abundances at
lower temperatures than VVEB, but none with showing the reverse. This is
consistent with an absence of depletion at higher temperatures, whilst the
sources at low temperature and high abundance could be explained by having
additional contributions of C$^{17}$O for regions outside the modelled core.
This would be consistent with the C$^{17}$O linewidths lying intermediate
between the two components of the C$^{18}$O lines as discussed above.
\linespread{1}
\linespread{1.4}

\linespread{1}
\begin{table*}

\begin{scriptsize}
  \begin{center}
 \begin{tabular}{cccccccccccc}

\toprule
      \toprule

     \multicolumn{1}{c}{}      &
     \multicolumn{4}{c}{C$^{17}$O \j{2}{1}}            &
     \multicolumn{6}{c}{C$^{18}$O \j{2}{1}}       \\
      \cmidrule(r){2-5}
     \cmidrule(l){6-11}

      \multicolumn{1}{c}{WFS}   &
      \multicolumn{1}{c}{$\Delta v$(Gauss)}             &
       \multicolumn{1}{c}{$v_{lsr}$}             &
      \multicolumn{1}{c}{$\Delta v$(hfs)}                &
         \multicolumn{1}{c}{$v_{lsr}$}             &
      \multicolumn{1}{c}{$\Delta v$(Gauss)}                &
        \multicolumn{1}{c}{$v_{lsr}$}             &
      \multicolumn{1}{c}{$\Delta v$(narrow)}     &
      \multicolumn{1}{c}{$v_{lsr}$}             &
      \multicolumn{1}{c}{$\Delta v$(broad)}     &    
         \multicolumn{1}{c}{$v_{lsr}$}             \\

  \midrule
14  & 2.84 &33.26 &2.44 &33.22 &2.42 & 33.26 &2.14 &33.21  &3.31 &33.47  \\
16  & 3.35 &59.21 &2.99 &59.19 &3.36 & 59.22 &2.43 &60.40  &2.46 &58.63 \\
17  & 3.02 &45.12 &2.65 &45.09 &2.72 & 45.08 &2.20 &44.87  &3.70 &45.67  \\
20  & 2.99 &34.35 &2.70 &34.29 &2.82 & 34.39 &1.60 &34.99  &2.97 &34.03  \\
21  & 2.76 &34.40 &2.25 &34.35 &2.25 & 34.39 &1.97 &34.46  &3.94 &33.61  \\
29  & 2.59 &59.28 &2.18 &59.28 &2.69 & 58.94 &1.83 &58.42  &1.93 &59.93 \\
30  & 3.05 &95.64 &2.96 &95.75 &2.44 & 95.69 &1.46 &95.65  &5.80 &95.79  \\
39  & 3.08 &96.07 &2.61 &96.03 &2.64 & 96.07 &1.55 &96.36  &3.72 &95.53  \\
78  & 2.82 &21.84 &2.47 &21.80 &2.83 & 21.72 &0.84 &21.70  &3.10 &21.73  \\
79  & 2.91 &22.60 &2.33 &22.54 &2.85 & 22.53 &1.30 &22.49  &3.96 &22.66  \\
90  & 3.25 & -3.63 &2.94 &-3.67 &2.83 &-3.64 &1.28 &-3.89  &3.59 &-3.41  \\
108 & 3.06 &-53.06 &2.87 &-53.10 &2.79 &-53.18 &1.50 &-53.06  &4.16 &-53.37  \\
109 & 2.56 &-44.35 &2.33 &-44.38 &2.74 &-44.29 &1.80 &-44.58  &3.57 &-43.82  \\
110 & 2.97 &-55.00 &2.09 &-54.74 &2.31 &-54.71 &1.35 &-54.66  &3.41 &-54.24  \\
112 & 2.03 &-18.53 &1.55 &-18.62 &1.84 &-18.45 &1.18 &-17.41 &1.31  &-18.65 \\
\hline

\end{tabular}
\end{center}
\end{scriptsize}
  \caption{\footnotesize All columns (except column 1) in units of km\,s$^{-1}$. The Gaussian linewidths were obtained using SPECX, whilst the hyperfine linewidths were   obtained using the hfs method in CLASS. Multiple Gaussian components are only given for those sources where appropriate.} 
\label{tab:lw}
\end{table*}


\linespread{1.4}

The degree of depletion in a region can provide an estimate of the
time for which gas has been cold and at high density. This provides a
measure of timescales for the formation of high-mass stars. 
Models of the physical and chemical changes occurring in the gas and
on dust grains during the pre-protostellar phase of the evolution of
cores, lead to predictions for the expected abundances of molecules
such as CO as a function of time \citep{viti99,flower05}.  With
detailed knowledge of depletion levels they become a viable tool for
determining the ages of the cores.

The detection of both C$^{17}$O and C$^{18}$O towards all of the sample shows
that that there are no sources towards which essentially all CO has been
depleted. This places an upper limit on the time for which the dust can have
existed at low temperatures of T $<$ 10\,K.  As we do not find depletion to
exceed more than a factor of $\sim$10 we can use the abundances predicted by
\citet{flower05} to estimate that the cores being traced are not older than
$\sim$3.5$\times 10^5$ years.  (For a depletion factor of 5 this reduces to
$\sim$2.2$\times 10^5$ years).

\subsection{Evidence of Source Evolution?}

\begin{figure}
\begin{center}
\includegraphics[angle=270,width=0.98\hsize]{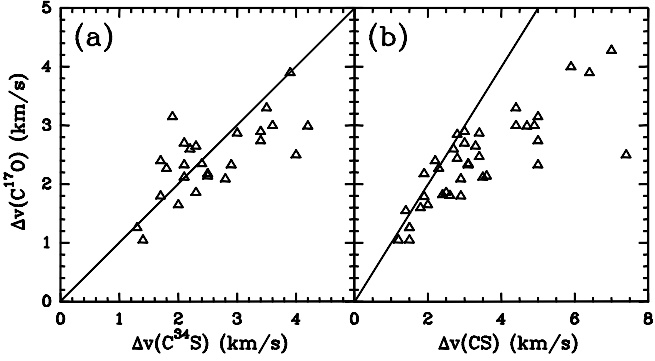}
\caption{{\footnotesize Plot showing the  C$^{17}$O hfs linewidth against
    {\bf(a)} the C$^{34}$S (\j{3}{2}) and {\bf(b)} the CS (\j{3}{2})
    linewidths \citep{beuther02a}. The solid lines show the
    unity ratio.}} 
\label{fig:cscolw}
\end{center}
\end{figure}

This set of sources has also been observed by \citet{beuther02a} at 1.2 mm
continuum emission and a number of CS lines. Figure \ref{fig:cscolw} shows a
comparison of the linewidth of C$^{17}$O \j{2}{1} (determined from the fit to
the hyperfine structure of the transition) against the linewidths of CS
\j{3}{2} and C$^{34}$S \j{3}{2}. It is clear that at smaller linewidths the
two are most closely correlated, with approximately equal linewidths. However
the linewidths tend to have a larger ratio at larger linewidths.   For CS
  linewidths $<3$~km\,s$^{-1}$ the ratio of the CS linewidth to the
  \cseventeeno\ linewidth is $1.17\pm0.05$, while for CS linewidths
  $\geq3$~km\,s$^{-1}$, the ratio is $1.56\pm0.09$. The same trend is also seen in
  the rarer species C$^{34}$S, where the ratio for C$^{34}$S linewidths
  $<3$~km\,s$^{-1}$ is $1.00\pm0.05$ and $1.22\pm0.07$ for the larger linewidth
  sources, suggesting this is not a result of high optical depth in the CS
  line.   We speculate that this increasing ratio of CS to \cseventeeno\
linewidth could arise as a result of CS being more affected than the
\cseventeeno\ by heating and stirring of the material by the central source,
through the action of the outflow.

\linespread{1}
\linespread{1.4}

\begin{figure}
\begin{center}
\includegraphics[angle=270,width=0.98\hsize]{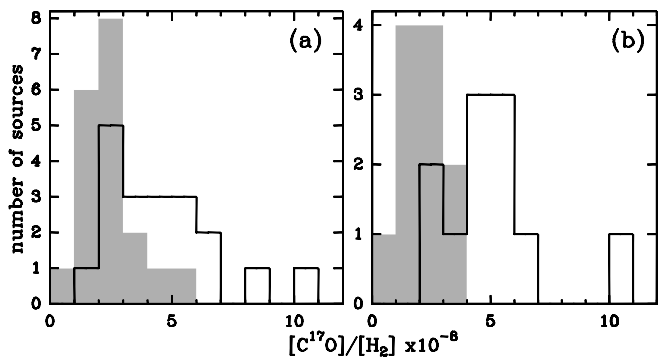}
\caption{\footnotesize {Distribution of \cseventeeno\ abundance. {\bf(a)} The
  solid bars show the distribution of abundances for those sources with
  $\Delta v$(CS) $>$ 3\,km\,s$^{-1}$ whilst the hollow bars represent those
  sources with $\Delta v$(CS) $<$ 3\,km\,s$^{-1}$. {\bf(b)} The solid bars show
  the distribution of abundances for the top 30\% of the sample ($\Delta v$(CS) $\gg$ 3\,km\,s$^{-1}$) while the hollow bars show the lower 30\% of the
  sample ($\Delta v$(CS) $\ll$ 3\,km\,s$^{-1}$).}}
\label{fig:abdhisto}
\end{center}
\end{figure}

 Interestingly there appears to be a connection between the \cseventeeno\
  abundance towards a source and its CS linewidth.  Comparing the
  distributions of C$^{17}$O abundance for those sources with $\Delta v$(CS)
  $>$ 3\,km\,s$^{-1}$ and those for which $\Delta v$(CS) $<$ 3\,km\,s$^{-1}$, a
  Kolmogorov-Smirnov two-sample test indicates a 98.6\% likelihood of a
  difference between the two groups (see Figure \ref{fig:abdhisto}a).  We
also considered just the upper ($\Delta v$(CS) $\gg$ 3\,km\,s$^{-1}$) $\sim$ 30\% of the
sample, comparing its distribution of \cseventeeno\ abundance with that of the
lower ($\Delta v$(CS) $\ll$ 3\,km\,s$^{-1}$) $\sim$ 30\% of the sample.
Repeating the K-S test for these groups of sources increases the statistical
significant of the difference between these new groups to 99.7\% (see Figure
\ref{fig:abdhisto}b), despite the smaller number of sources in each group.
This increase in the difference between the groups might be expected if these
groups are not distinct categories but rather represent two extremes in a
continuum of source properties, from small linewidth, high \cseventeeno\
abundance objects to large linewidth, low \cseventeeno\ abundance objects,
perhaps tracing an evolutionary progression.

To further probe the statistical significance of this difference between the
sources we performed two tests.  In the first we randomly shuffled the
\cseventeeno\ abundances amongst  the sources. The sample was then again
  divided between objects with $\Delta v$(CS) $>$ 3\,km\,s$^{-1}$ and those for
which $\Delta v$(CS) $<$ 3\,km\,s$^{-1}$.   A K-S test was then performed
  comparing the resulting distributions of \cseventeeno\ abundance for the two
  linewidth groups.  Repeating this randomisation and testing 100 times we
found no cases where the statistical difference between the shuffled
abundances exceed that of actual data.   In other words the difference in
  \cseventeeno\ abundance between the sources with large linewidths and those
  with small linewidths has a probability of less than 1\% of being a chance
  coincidence.

For the second test we take into account the observational uncertainties
associated with the abundances by randomly resampling the abundances
associated with each source.  For each source we regenerated a \cseventeeno\
abundance by randomly drawing an abundance from a Gaussian distribution with a
dispersion equal to the uncertainty in the measured \cseventeeno\ abundance of
that source.  We did this for the whole sample, split the sample on the basis
of the CS linewidth and again apply a K-S test to intercompare the resulting
distributions of \cseventeeno\ abundance.  This was repeated 100 times. We
found that on average the two samples  differed at 98\% confidence level,
  with the confidence level never less than 97\%. This indicates that the
  difference in abundance between our two groups is unlikely to arise as a
  result of the statistical uncertainties.

 The association of low \cseventeeno\ abundance with sources with
  large linewidths is somewhat surprising. However it can be
  understood if the low \cseventeeno\ abundance (which is calculated
  integrated across the whole line profile and averaged across the
  telescope beam) reflects the conditions in the cold environment in the
  outer, extended envelope around the central source, while the broad
  linewidth reflects some component close to the central source,
  possibly related to outflow. Mapping the spatial distribution
  of the depletion towards these objects by comparing molecular line
  and deep dust continuum maps would be able to test this
  interpretation.

  Whilst the central regions of these regions certainly have temperatures
  reflecting the presence of a luminous embedded source, the models of
  WFS05 imply that the majority of the extended envelope lies
  at temperatures below 50\,K. In many cases the models show the temperature
  dropping below 20\,K towards the outer regions of the envelopes, temperatures
  which are similar to those derived from the NH$_3$ observations of SBSMW.
  It is presumably in these outer regions, far from the heating and effects of
  the outflow from the central source, that the \cseventeeno\ is frozen out
  onto the surface of dust grains.  

The high \cseventeeno\ abundances displayed by those sources with the lowest
linewidths could be consistent with two possible scenarios; either these
sources are extremely young and depletion has not yet occurred to a
significant degree, or alternatively the sources are much older having
undergone substantial heating by an embedded population of sources, or
  shocks from outflows, which has evaporated the ice mantles containing the
  \cseventeeno\ back into the gas phase.

 This second scenario could be confirmed by the positive detection towards these sources of the molecules which trace hot core emission.  SBSMW have searched many of these sources for
  such species and shown that at least some of these sources are
  evolved enough to have locally heated their natal cores to
  temperatures of $>$100\,K.  However the failure to detect these
  molecules towards some sources implies that these sources are either
  highly evolved with the hot core molecules have been destroyed by
  ion reactions and ages $\sim$10$^5$ years (e.g.  Hatchell et al.
    1998), or else would suggest that these are extremely young
  sources.  A more detailed discussion of these possible scenarios is
presented in \citet{tf}.  High spatial resolution observations of
these regions would help to distinguish between the two pictures by
  identifying the location of the depleted material.  Very young cores
  would be expected to be depleted in \cseventeeno\ in their centres
  where the density is highest but to have normal \cseventeeno\
  abundance in their more extended, lower density, envelopes. For a
  more evolved core which has formed a high mass protostar at its
  centre, this central object will be heating the interior region of
  the core, releasing frozen out \cseventeeno\ back into the gas
  phase. However sufficient time may have passed for \cseventeeno\ to
  be depleted in the outer parts of the core.

\linespread{1}
\linespread{1.4}

\section{Summary}

We have presented spectra of 84 candidate HMPOs in the \j{2}{1} line
of C$^{17}$O and for a selection of these we have also observed
C$^{18}$O \j{2}{1} and C$^{17}$O \j{3}{2}. We have used the data to
calculate the \cseventeeno\ column densities and for those sources for
which models of dust emission exist we constructed models of the CO
emission to check the validity of our assumptions regarding excitation
temperature and dust temperature. We have found the following:

The ratio of C$^{17}$O column density to H$_2$ derived from 850\,$\mu$m
dust emission has a significant scatter across our sample. The scatter
between the most extreme sources is $\sim$14 and is consistent with
the abundance of C$^{17}$O differing on a source-by-source basis.

Our derivation of the H$_2$ column density assumed a constant value for
$\kappa$ for all the sources, however there are no corrections to $\kappa$
which when applied globally to the sample to eliminate this scatter. Indeed
the elimination of the scatter would require the sources lying at the extremes
of the scatter to possess dust properties which are physical unlikely.

The method and assumptions used to calculate the \cseventeeno\ column
densities were tested by modelling a subset of these sources with the
radiative transfer code RATRAN. For all suitable models the best-fit abundance
of C$^{17}$O matched the data to within a factor of $\sim$2, supporting our
assumptions and removing the possibility of the assumptions accounting for the
range of abundances.  We conclude that this scatter arises as a result of
depletion present in a number of these sources.  However the maximum degree of
depletion we find does not exceed $\sim$7. This range of depletion suggests
that  outer regions of these sources have lifetimes  during which
  they are cold and dense of about $2-4\times10^5$ years.

We derive the mass-weighted temperature and find a correlation with C$^{17}$O
abundance, consistent with VVEB who from this correlation also infer
the presence of depletion towards sources similar to those discussed here.

These sources have been studied by \citet{beuther02a} and a comparison of our
C$^{17}$O linewidths to their CS \j{3}{2} linewidths shows a division between
those which have $\Delta v$(CS) $>$ 3\,km\,s$^{-1}$ and those for which $\Delta
v$(CS) $<$ 3\,km\,s$^{-1}$. A statistical comparison of the \cseventeeno\
abundances of these two groups reveal a trend of the first group displaying
significantly lower abundances than those in the second group with a
statistical significance exceeding $\sim$98\%.

We suggest that this range of \cseventeeno\ abundance and the implied degree
of depletion between the sources may reflect a spread in evolutionary status
amongst the sources.  Higher angular resolution observations of dust emission,
\cseventeeno\ and other species towards these objects will be important in
confirming this suggestion.  It is possible that high resolution
observations will identify regions with higher degrees of depletion than can
be measured with the relatively large telescope beam used for the observations
presented here.

Depletion towards these objects shows that during the evolution of these
  cores the gas has remained cold and dense for long enough for the trace
species to deplete. More detailed models of how trace molecules deplete in
massive dense cores, combined with higher angular resolution observations,
should be able to provide tighter constraints on the lifetime of these
regions, including how long they exist before they form massive stars.

\linespread{1}

\begin{table*}
\begin{footnotesize}
  \begin{center}
    \begin{tabular}{llccccc}
      \toprule
      \toprule

      \multicolumn{1}{l}{WFS}            &
      \multicolumn{1}{c}{IRAS NAME}           &
      \multicolumn{1}{c}{R.A.}      &
      \multicolumn{1}{c}{Dec.}      &
      \multicolumn{3}{c}{DATES OF OBSERVATIONS}    \\

      \cmidrule(r){3-4}\cmidrule(l){5-7}

      \multicolumn{1}{c}{}                        &
      \multicolumn{1}{c}{}                      &
      \multicolumn{1}{c}{(J2000)}              &
      \multicolumn{1}{c}{(J2000)}              &
      \multicolumn{1}{c}{C$^{18}$O\,(2$-$1)}                     &
      \multicolumn{1}{c}{C$^{17}$O\,(2$-$1)} &
      \multicolumn{1}{c}{C$^{17}$O\,(3$-$2)} \\     
     
      \midrule

\object{WFS1}  &\object{IRAS 05358+3543}  &05 39 10.8   &$+$35 45 16   &  &05/07/21 &05/07/23    \\
\object{WFS3}  &\object{IRAS 05490+2658}  &05 52 11.0   &$+$27 00 34   &  &05/07/21 &05/07/23    \\
\object{WFS4}  &\object{IRAS 05490+2658}  &05 52 12.1   &$+$27 00 11   &  &05/07/21 &05/07/23    \\
\object{WFS6}  &\object{IRAS 05553+1631}  &05 58 13.4   &$+$16 32 00   &  &05/07/21 &05/07/23    \\
\object{WFS7}  &\object{IRAS 18089-1732}  &18 11 45.2   &$-$17 30 43   &  & &05/07/07    \\
\object{WFS8}  &\object{IRAS 18089-1732}  &18 11 51.5   &$-$17 31 34   &  & &05/07/07    \\
\object{WFS11} &\object{IRAS 18089-1732}  &18 11 57.0   &$-$17 29 34   &  & &05/07/07    \\
\object{WFS12} &\object{IRAS 18090-1832}  &18 12 02.1   &$-$18 31 58   &  &05/08/03 &    \\
\object{WFS13} &\object{IRAS 18102-1800}  &18 13 11.7   &$-$18 00 04   &  &05/08/03 &    \\
\object{WFS14} &\object{IRAS 18151-1208}  &18 17 58.2   &$-$12 07 28   &04/05/13 &04/05/13, 04/05/14&    \\
\object{WFS15} &\object{IRAS 18159-1550}  &18 18 48.4   &$-$15 49 00   &  &05/08/03 &    \\
\object{WFS16} &\object{IRAS 18182-1433}  &18 21 08.9   &$-$14 31 46   &04/05/13  &04/05/13  &     \\
\object{WFS17} &\object{IRAS 18223-1243}  &18 25 10.6   &$-$12 42 27  &04/05/13  &04/05/14 &    \\
\object{WFS18} &\object{IRAS 18247-1147}  &18 27 31.4   &$-$11 45 55   &  &05/07/27, 05/08/03 &    \\
\object{WFS19} &\object{IRAS 18264-1152}  &18 29 14.3   &$-$11 50 22   &  &05/07/27, 05/08/03 &    \\
\object{WFS20} &\object{IRAS 18272-1217}  &18 30 02.2   &$-$12 15 40  &04/05/13  &04/05/14 &    \\
\object{WFS21} &\object{IRAS 18272-1217}  &18 30 03.2   &$-$12 15 11  &04/05/13  &04/05/14 &    \\
\object{WFS22} &\object{IRAS 18290-0924}  &18 31 43.4   &$-$09 22 26   &04/05/14  &04/05/14 &05/07/22    \\
\object{WFS23} &\object{IRAS 18290-0924}  &18 31 44.0   &$-$09 22 17   &  &05/08/03 &    \\
\object{WFS24} &\object{IRAS 18306-0835}  &18 33 17.3   &$-$08 33 28   &  &05/08/03 &    \\
\object{WFS25} &\object{IRAS 18306-0835}  &18 33 23.9   &$-$08 33 33  &04/05/13  &04/05/13 &    \\
\object{WFS28} &\object{IRAS 18310-0825}  &18 33 47.9   &$-$08 23 52   &  &05/08/03 &    \\
\object{WFS29} &\object{IRAS 18337-0743}  &18 36 27.9   &$-$07 40 25   &04/05/14  &04/05/14 &05/07/22    \\
\object{WFS30} &\object{IRAS 18345-0641}  &18 37 16.8   &$-$06 38 35  &04/05/13  &04/05/13 &    \\
\object{WFS33} &\object{IRAS 18348-0616}  &18 37 30.5   &$-$06 14 13   &  &05/07/27, 05/08/03 &    \\
\object{WFS34} &\object{IRAS 18372-0541}  &18 37 16.8   &$-$06 38 35  &04/05/14, 04/05/15  &04/05/14 &    \\
\object{WFS35} &\object{IRAS 18385-0512}  &18 41 12.8   &$-$05 08 58   &  &05/08/03 &    \\
\object{WFS36} &\object{IRAS 18426-0204}  &18 45 12.1   &$-$02 01 10  &04/05/14, 04/05/15  &04/05/14 &    \\
\object{WFS37} &\object{IRAS 18431-0312}  &18 45 45.5   &$-$03 09 21   &  &05/07/27, 05/08/03 &    \\
\object{WFS38} &\object{IRAS 18437-0216}  &18 46 21.8   &$-$02 12 20   &  &05/07/27, 05/08/03 &    \\
\object{WFS39} &\object{IRAS 18437-0216}  &18 46 22.4   &$-$02 14 16   &04/05/14, 04/05/15  &04/05/14, 04/05/15 &05/07/07    \\
\object{WFS41} &\object{IRAS 18440-0148}  &18 46 33.3   &$-$01 44 52   &  &05/08/03 &    \\
\object{WFS42} &\object{IRAS 18440-0148}  &18 46 36.5   &$-$01 45 22   &  &05/08/03 &    \\
\object{WFS51} &\object{IRAS 18460-0307}   &18 48 37.8  &$-$03 03 48   &04/05/14, 04/05/15  &04/05/14, 04/05/15 &05/07/07    \\
\object{WFS55} &\object{IRAS 18472-0022}   &18 49 52.4  &$-$00 18 59   &  &05/08/03 &    \\
\object{WFS57} &\object{IRAS 18488+0000}   &18 51 24.4  &$+$00 04 39   &  &05/07/26 &    \\
\object{WFS58} &\object{IRAS 18488+0000}   &18 51 25.5  &$+$00 04 11   &  &05/07/26 &    \\
\object{WFS59} &\object{IRAS 18521+0134}   &18 54 40.6  &$+$01 38 05   &  &05/08/03 &    \\
\object{WFS60} &\object{IRAS 18521+0134}   &18 54 44.4  &$+$01 37 00   &  &05/08/03 &    \\
\object{WFS61} &\object{IRAS 18530+0215}   &18 55 33.7  &$+$02 19 09   &  &05/08/03 &    \\
\object{WFS62} &\object{IRAS 18540+0220}   &18 56 36.6  &$+$02 24 45   &  &05/07/30 &    \\
\object{WFS63} &\object{IRAS 18540+0220}   &18 56 40.1  &$+$02 25 30   &  &05/07/30 &    \\
\object{WFS64} &\object{IRAS 18553+0414}   &18 57 53.5  &$+$04 18 16   &  &05/07/30 &    \\
\object{WFS66} &\object{IRAS 19012+0536}   &19 03 45.3  &$+$05 40 43   &  &05/08/03 &    \\
\object{WFS67} &\object{IRAS 19035+0641}   &19 06 01.5  &$+$06 46 35   &  &05/08/03 &    \\
\object{WFS68} &\object{IRAS 19074+0752}   &19 09 53.4  &$+$07 57 12   &  &05/07/29 &    \\
\object{WFS69} &\object{IRAS 19074+0752}   &19 09 53.9  &$+$07 56 55   &  &05/07/29 &    \\
\object{WFS70} &\object{IRAS 19175+1357}   &19 19 48.6  &$+$14 02 26   &  &05/07/27 &    \\
\object{WFS71} &\object{IRAS 19175+1357}   &19 19 48.8  &$+$14 02 46   &  &05/07/26 &    \\
\object{WFS72} &\object{IRAS 19217+1651}   &19 23 58.6  &$+$16 57 38   &  &05/07/27 &    \\
\object{WFS74} &\object{IRAS 19266+1745}   &19 28 55.5  &$+$17 52 00   &  &05/07/30 &    \\
\object{WFS75} &\object{IRAS 19282+1814}   &19 30 23.1  &$+$18 20 22   &  &05/07/30 &    \\
\object{WFS76} &\object{IRAS 19282+1814}   &19 30 29.7  &$+$18 20 37   &  &05/07/30 &    \\
\object{WFS77} &\object{IRAS 19403+2258}   &19 42 28.8  &$+$23 05 03   &  &05/07/28 &05/07/07    \\
\object{WFS78} &\object{IRAS 19410+2336}   &19 43 10.6  &$+$23 45 02   &04/05/14  &04/05/14 &05/07/07    \\
 \midrule
      \multicolumn{7}{r}{\small\sl continued on next page}\\
      \midrule

\end{tabular}
\end{center}

\end{footnotesize}
\end{table*}

\begin{table*}
\begin{footnotesize}
  \begin{center}

    \begin{tabular}{llccccc}

     \toprule
     \multicolumn{5}{l}{\small\sl continued from previous page} \\
   \toprule
      \multicolumn{1}{l}{WFS}            &
      \multicolumn{1}{c}{IRAS NAME}           &
      \multicolumn{1}{c}{R.A.}      &
      \multicolumn{1}{c}{Dec.}      &
      \multicolumn{3}{c}{DATES OF OBSERVATIONS}    \\

      \cmidrule(r){3-4}\cmidrule(l){5-7}

      \multicolumn{1}{c}{}                        &
      \multicolumn{1}{c}{}                      &
      \multicolumn{1}{c}{(J2000)}              &
      \multicolumn{1}{c}{(J2000)}              &
      \multicolumn{1}{c}{C$^{18}$O\,(2$-$1)}                     &
      \multicolumn{1}{c}{C$^{17}$O\,(2$-$1)} &
      \multicolumn{1}{c}{C$^{17}$O\,(3$-$2)} \\
     
      \midrule

\object{WFS79}  &\object{IRAS 19410+2336}   &19 43 11.2  &$+$23 44 06   &04/05/14  &04/05/14, 04/05/23  &05/07/07    \\
\object{WFS80}  &\object{IRAS 19411+2306}   &19 43 17.6  &$+$23 13 57   &  &05/07/28 &05/07/07    \\
\object{WFS81}  &\object{IRAS 19413+2332}   &19 43 26.3  &$+$23 40 26   &  &05/07/28 &05/07/07    \\
\object{WFS82}  &\object{IRAS 19413+2332}   &19 43 29.0  &$+$23 40 19   &  &05/07/29 &05/07/07    \\
\object{WFS83}  &\object{IRAS 19471+2641}   &19 49 10.1  &$+$26 49 10   &  &05/07/29 &05/07/07    \\
\object{WFS84}  &\object{IRAS 19471+2641}   &19 49 11.8  &$+$26 49 38   &  &05/07/29 &05/07/07    \\
\object{WFS85}  &\object{IRAS 20051+3435}   &20 07 04.5  &$+$34 44 45  &  &04/05/23 &    \\
\object{WFS86}  &\object{IRAS 20081+2720}   &20 10 12.6  &$+$27 29 13   &  &05/07/26 &    \\
\object{WFS87}  &\object{IRAS 20081+2720}   &20 10 13.3  &$+$27 28 21  &  &04/05/23 &    \\
\object{WFS88}  &\object{IRAS 20081+2720}   &20 10 16.0  &$+$27 28 12  &  &04/05/23 &    \\
\object{WFS89}  &\object{IRAS 20081+2720}   &20 10 18.7  &$+$27 27 18   &  &05/07/26 &    \\
\object{WFS90}  &\object{IRAS 20126+4104}   &20 14 25.7  &$+$41 13 30   &04/05/14  &04/05/13 &05/07/21    \\
\object{WFS91}  &\object{IRAS 20205+3948}   &20 22 20.0  &$+$39 58 21   &04/05/14  &04/05/13 &05/07/21    \\
\object{WFS92}  &\object{IRAS 20205+3948}   &20 22 24.9  &$+$39 57 55   &  &05/07/26 &    \\
\object{WFS93}  &\object{IRAS 20216+4107}   &20 23 23.9  &$+$41 17 42   &  &05/07/26 &    \\
\object{WFS94}  &\object{IRAS 20293+3952}   &20 31 12.9  &$+$40 03 21   &  &05/07/27 &    \\
\object{WFS95}  &\object{IRAS 20319+3958}   &20 33 49.4  &$+$40 08 32  &  &04/05/15 &    \\
\object{WFS96}  &\object{IRAS 20332+4124}   &20 34 58.7  &$+$41 34 46  &  &04/05/15 &    \\
\object{WFS97}  &\object{IRAS 20332+4124}   &20 35 01.1  &$+$41 34 59  &  &04/05/15 &    \\
\object{WFS98}  &\object{IRAS 20343+4129}   &20 36 03.4  &$+$41 39 44   &  &05/07/27 &    \\
\object{WFS99}  &\object{IRAS 20343+4129}   &20 36 06.3  &$+$41 39 59  &        04/05/17  &04/05/23 &    \\
\object{WFS100} &\object{IRAS 20343+4129}  &20 36 08.1  &$+$41 39 58  &        04/05/17  &04/05/23 &    \\
\object{WFS101} &\object{IRAS 22134+5834}  &22 15 08.9   &$+$58 49 08   &  &05/07/23 &    \\
\object{WFS102} &\object{IRAS 22551+6221}  &22 57 04.3   &$+$62 37 44   &  &05/07/23 &    \\
\object{WFS103} &\object{IRAS 22551+6221}  &22 57 07.4   &$+$62 37 29   &  &05/07/23 &    \\
\object{WFS104} &\object{IRAS 22551+6221}  &22 57 11.6   &$+$62 36 46   &  &05/07/23 &    \\
\object{WFS107} &\object{IRAS 22570+5912}  &22 59 05.0   &$+$59 28 23   &04/05/13  &04/05/13 &05/07/26    \\
\object{WFS108} &\object{IRAS 23033+5951}  &23 05 24.8   &$+$60 08 14   &04/05/13  &04/05/13 &05/07/26    \\
\object{WFS109} &\object{IRAS 23139+5939}  &23 16 09.8   &$+$59 55 31   &04/05/13  &04/05/13 &05/07/26    \\
\object{WFS110} &\object{IRAS 23151+5912}  &23 17 20.4   &$+$59 28 51   &04/05/14  &04/05/22 &05/07/26    \\
\object{WFS111} &\object{IRAS 23545+6508}  &23 57 02.1   &$+$65 24 38   &04/05/14  &04/05/22 &05/07/26    \\
\object{WFS112} &\object{IRAS 23545+6508}  &23 57 06.4   &$+$65 24 49   &04/05/14  &04/05/22 &05/07/26    \\

\bottomrule     

\end{tabular}
\end{center}

\caption{ The telescope pointing coordinates are given
  along with the associated IRAS source found within the frame. The
  WFS labels refer to the submm peaks identified in \citet{wfs04}.}
\label{tab:sourcelist}

\end{footnotesize}

\end{table*}

\linespread{1.4}
\bibliographystyle{aa}

\end{document}